\documentclass[a4paper,12pt]{article}
\usepackage{amsmath,amssymb,amsthm}
\usepackage[totalwidth=16.5cm,totalheight=23.5cm]{geometry}
\usepackage{graphicx}
\newcommand{\R}{{\mathbb R}}
\newcommand{\C}{{\mathbb C}}

\newcommand{\T}{{\mathcal T}}

\newcommand{\bbS}{{\mathbb S}}

\newcommand{\im}{{\rm i}}
\newcommand{\tr}{{\rm tr}}
\newcommand{\Id}{{\rm Id}}
\newcommand{\Diff}{{\rm Diff}}
\newcommand{\QS}{{\rm QS}}
\newcommand{\II}{I\kern -0.7ex I}
\newcommand{\III}{I\kern -0.7ex I\kern -0.7ex I}

\newcommand{\be}{\begin{eqnarray}}
\newcommand{\ee}{\end{eqnarray}}

\begin{document}
\pagestyle{plain}
\title{The universal phase space of AdS${}_3$ gravity}
\author{Carlos Scarinci${}^1$ and Kirill Krasnov${}^{1,2}$ \\ \\ \it{${}^1$ \, School of Mathematical Sciences, University of Nottingham}\\ \it{University Park, Nottingham, NG7 2RD, UK} \and 
\it{${}^2$ \, Max Planck Institute for Gravitational Physics}  \\ \it{Am M\"uhlenberg 1, 14476 Golm, Germany}}
\date{}
\maketitle
\begin{abstract}\noindent
We describe what can be called the ``universal'' phase space of AdS${}_3$ gravity, in which the moduli spaces of globally hyperbolic AdS spacetimes with compact
 spatial sections, as well as the moduli spaces of multi-black-hole spacetimes are realized as submanifolds. The universal phase space is parametrized by two
 copies of the universal Teichm\"uller space $\T(1)$ and is obtained from the correspondence between maximal surfaces in AdS${}_3$ and quasisymmetric homeomorphisms
 of the unit circle. We also relate our parametrization to the Chern-Simons formulation of 2+1 gravity and, infinitesimally, to the holographic (Fefferman-Graham)
 description. In particular, we obtain a relation between the generators of quasiconformal deformations in each $\T(1)$ sector and the chiral Brown-Henneaux vector
 fields. We also relate the charges arising in the holographic description (such as the mass and angular momentum of an AdS${}_3$ spacetime) to the periods of the quadratic
 differentials arising via the Bers embedding of $\T(1)\times\T(1)$. Our construction also yields a symplectic map $T^\ast\T(1)\rightarrow\T(1)\times\T(1)$ generalizing
 the well-known Mess map in the compact spatial surface setting.
\end{abstract}
\section{Introduction}

Since the discoveries by Brown and Henneaux \cite{Brown:1986nw} that the group of symmetries of an asymptotically AdS${}_3$ spacetime is a centrally extended conformal
 group in two dimensions, and then by Banados, Teitelboim and Zanelli \cite{Banados:1992wn} that black holes can exist in such spacetimes, the subject of negative
 cosmological constant gravity in 2+1 dimensions continues to fascinate researchers. The result \cite{Brown:1986nw}  is now considered to be a precursor of the
 AdS/CFT correspondence of string theory \cite{Witten:1998qj}, and the value of the central charge determined in \cite{Brown:1986nw} is an essential ingredient
 of the conformal field theoretic explanation \cite{Strominger:1997eq} of the microscopic origin of the black hole entropy.

The AdS${}_3$/CFT${}_2$ story is reasonably well-understood in the string theory setting of 3-dimensional gravity coupled to a large number of fields of string
 (and extra dimensional) origin. At the same time, the question of whether there is a CFT dual to pure AdS${}_3$ gravity remains open, see \cite{Witten:2007kt}
 and \cite{Maloney:2007ud} for the most recent (unsuccessful) attempts in this direction. In particular, the attempt  \cite{Maloney:2007ud} to construct the
 genus one would-be CFT partition function by summing over the modular images of the partition function of pure AdS leads to discouraging conclusions. It thus
 appears that pure AdS${}_3$ gravity either does not have enough ``states'' to account for the BH entropy microscopically, or that the known such states cannot
 be consistently put together into some CFT structure.

The current lack of understanding of pure AdS${}_3$ gravity quantum mechanically is particularly surprising given the fact that, in a sense, the theory is trivial
 since pure gravity in 2+1 dimensions does not have any propagating degrees of freedom. In the setting of compact spatial sections the phase space of 2+1 gravity
 (i.e. the space of constant curvature metrics in a $\R\times\Sigma$, with $\Sigma$ a genus $g>1$ Riemann surface) is easy to describe (for all values of the
 cosmological constant). The constant mean curvature foliation of such a spacetime is particularly useful for this purpose. One finds, see \cite{Moncrief:1989dx}
 and also \cite{Krasnov:2005dm} for a more recent description, that the phase space is the cotangent bundle over the Teichm\"uller space of the spatial slice
 (for any value of the cosmological constant). The zero cosmological constant result \cite{Moncrief:1989dx} also follows quite straightforwardly from the Chern-Simons
 (CS) description given in \cite{Witten:1988hc}. In the setting of AdS${}_3$ manifolds with compact spatial slices, there is yet another description of the same phase
 space, first discovered by Mess \cite{Mess:2007}. This is given by two copies of the Teichm\"uller space of the spatial slice, or, equivalently, by two hyperbolic
 metrics on the spatial slice Riemann surface. The Mess description is related to the Chern-Simons description of AdS${}_3$ gravity in terms of two copies of
 ${\rm SL}(2,\R)$ CS theory. 

It appears sensible to tackle the problem of quantum gravity as a problem of quantization of the arising classical phase space. One could argue that this approach
 is unlikely to succeed in 3+1 and higher dimensions, where the phase spaces that arise this way are infinite dimensional (because of the existence of local
 excitations --- gravitational waves). However, in the setting of 2+1 gravity, at least in the setting of spacetimes with compact spatial slices one deals with
 a finite-dimensional dynamical system and the problem of quantum gravity seems to reduce to a problem from quantum mechanics. In spite of this being a tractable
 problem, the immediate worry with this approach is that the Hilbert space of quantum states one can obtain by quantizing such a finite-dimensional phase space
 would not be sufficiently large to account for the black hole entropy. 

At the same time, in the context of black holes one should consider non-compact spatial slices. The classical phase space that should arise in this context is
 somewhat less understood. On one hand, we now know that there is not just the simple BTZ BH \cite{Banados:1992wn}, but also a much more involved zoo of multi-black
 hole (MBH) spacetimes first described in \cite{Brill:1995jv}. A rather general description of such MBH's using causal diamonds at their conformal infinity is given
 in \cite{Barbot:2005qk}. As a by-product of a construction in  \cite{Bonsante:2006tr} using earthquakes, another description of MBH geometries is also available.
 There are also descriptions of MBH spacetimes in the physics literature, see \cite{Krasnov:2001va} and \cite{Skenderis:2009ju}. These descriptions show that, like
 in the compact spatial slice setting with its Mess parametrization, the geometry of multi-black-holes continues to be parametrized by two hyperbolic metrics on their
 spatial slice (or, equivalently, by the cotangent bundle of the corresponding Teichm\"uller space). The main difference with the compact setting is that the spatial slices
 are now Riemann surfaces with a geodesic boundary (or with hyperbolic ends attached), and there are now additional moduli, namely the sizes of the boundary components. These
 new length parameters, two for each boundary component (because there are two hyperbolic metrics involved in the parametrization) determine the geometrical characteristics of
 the corresponding black hole horizon, such as its length and angular velocity. An explicit formula of this sort can be found in e.g.  \cite{Bonsante:2006tr}, see formula (1) of
 the first (arxiv) version of this paper. All in all, there is a reasonable understanding of the geometry of the multi-black-hole spacetimes, as well as an efficient parametrization
 of these spacetimes by two copies of the Teichm\"uller space of Riemann surfaces with boundaries. It thus might seem that the phase space of non-compact spatial slices
 AdS${}_3$ gravity is as finite dimensional as in the compact setting.

It is however clear that a geometrical description of the multi-black-hole spacetime is just half of the story. Indeed, one of the most exciting aspects of 2+1 gravity
 in asymptotically AdS setting is the fact \cite{Brown:1986nw} that the diffeomorphisms that are asymptotically non-trivial should no longer be interpreted as gauge.
 Indeed, they map one asymptotically AdS spacetime into a non-equivalent one. Thus, asymptotic symmetries applied e.g. to the AdS${}_3$ create an infinitely large
 class of asymptotically AdS${}_3$ spacetimes described by Brown and Henneaux \cite{Brown:1986nw}. The phase space consisting of all such deformations of a reference
 spacetime is then infinite dimensional and the problem of its quantization therefore becomes much more non-trivial than in the compact spatial slice setting. It could
 be that the CFT dual of pure 2+1 gravity can be discovered by quantizing this phase space. And indeed, the states obtained from the AdS ``vacuum'' by an action of the
 Virasoro generators is what was summed over in \cite{Maloney:2007ud} in the authors' attempt to build the genus one pure gravity partition function. 

We can now formulate the main objective of this paper. Our main aim is to give a description of the phase space of AdS${}_3$ gravity that is equally applicable to both
 compact and non-compact spatial section spacetimes. At the same time, we would like our description to include the Brown-Henneaux asymptotic deformations. As we shall see,
 there is a ``universal'' way of doing so, where one constructs what can be called the universal phase space, in which all the moduli spaces of fixed spatial topology are
 realized as submanifolds. We achieve this in the same way as in the context of the universal Teichm\"uller space, where the fixed topology Teichm\"uller spaces are realized
 as (complex) submanifolds of the universal Teichm\"uller space. The construction of this paper can then be seen as a generalization of the description of \cite{Mess:2007} to
 the setting of the universal Teichm\"uller space.

We want to emphasize that we do not consider here the quantum theory that would arise by quantizing the classical phase space of 2+1 gravity. This is left to future
 studies. Rather, our main aim here is to describe the phase space in as explicit terms as possible, thus setting the stage for its quantization. We shall see that the
 universal phase space is extremely non-trivial, and is parametrized by two copies $\T(1)\times \T(1)$ of the universal Teichm\"uller space $\T(1)$, or, equivalently,
 the cotangent bundle $T^\ast\T(1)$ over $\T(1)$. This generalizes Mess's description \cite{Mess:2007} of the compact spatial slice setting, where there is similarly
 two equivalent parametrizations of the moduli space of AdS$_3$ spacetimes. 

Let us briefly indicate how the universal Teichm\"uller space comes about. In one possible definition of the latter, this is the space of quasisymmetric homeomorphisms of
 the unit circle (such homeomorphisms are boundary values of quasiconformal maps from the unit disc to itself). Thus, in very general terms, the universal phase space of
 AdS${}_3$ gravity is parametrized by two functions on the circle. To obtain the moduli space of spacetimes of fixed spatial topology, e.g. that of fixed topological type
 multi-black-holes, one imposes the condition that the functions in question be invariant under a suitable discrete subgroup of the group of M\"obius transformations. In the
 case of multi-black-hole spacetimes this produces a moduli space that is still infinite-dimensional and that includes the Brown-Henneaux ``excitations'' in all asymptotic
 regions. The cardinality is that of a pair of functions for each asymptotic region. While the freedom of prescribing two functions on the circle could be anticipated
 already from the Fefferman-Graham type description, see below, one novelty of our construction is that the phase space includes all possible multi-black-hole spacetimes.
 Another novelty of the constructions of this paper is a precise characterization of which functions on the circle are relevant in the context of AdS${}_3$ gravity. Indeed,
 our description in terms of two points in the universal Teichm\"uller space shows that these are associated to quasisymmetric maps of the unit circle which is larger than
 the class of smooth maps.

As we have already mentioned, the fact that in the non-compact spatial slice setting the phase space becomes an infinite dimensional space of certain (pairs of)
 functions on $\bbS^1$ can be expected already from the AdS/CFT perspective. Indeed, we know that a possible description of an asymptotically AdS spacetime is in
 terms of an expansion of the spacetime metric in a neighbourhood of the conformal boundary, see  e.g. \cite{Banados:1998,Skenderis:1999} for such expansions in the
 AdS${}_3$ context. For any asymptotically AdS${}_3$ spacetime one can find the so-called Fefferman-Graham coordinates in a neighbourhood of (a component of) the
 conformal boundary where the bulk metric takes the form
\be\nonumber
ds^2=\frac{d\rho^2}{\rho^2}+\frac{1}{\rho^2}(g_{(0)}+\rho^2 g_{(2)}+\rho^4 g_{(4)})
\ee
Here $g_{(0)}$ is a representative of the conformal class on the conformal boundary and
\be\nonumber
g_{(2)}=\frac{1}{2}\left(R_{(0)}g_{(0)}+T\right),\quad g_{(4)}=\frac{1}{4}g_{(2)}g_{(0)}^{-1}g_{(2)}
\ee
with $R_{(0)}$ the Ricci scalar of $g_{(0)}$ and $T$ the quasilocal stress-tensor \cite{Henningson:1998,Balasubramanian:1999}. Note that for fixed $g_{(0)}$ the
 only freedom in specifying the space time metric are the components of $T$.

For a flat boundary metric (which is always achievable by choosing $\rho$ appropriately), the most general quasilocal stress tensor can be written
\be\nonumber
T=adt^2+2bdtd\theta+ad\theta^2,
\ee
with $a$ and $b$ given by sum and difference of two arbitrary chiral functions
\be\nonumber
a(t,\theta)=a_+(t+\theta)+a_-(t-\theta),\quad b(t,\theta)=a_+(t+\theta)-a_-(t-\theta),
\ee
and the spacetime metric becomes
\begin{align}\label{intr-FH}
ds^2=&\;\frac{d\rho^2}{\rho^2}+\frac{1}{4\rho^2}(-dt^2+d\theta^2)+\frac{1}{2}(adt^2+2bdtd\theta+ad\theta^2)
\cr
& +\frac{\rho^2}{4}(a^2-b^2)(-dt^2+d\theta^2).
\end{align}
Note that the possibility of writing down the Fefferman-Graham type expansion in a closed form (with a finite number of terms) is peculiar to 2+1 dimensions,
 and is due to the absence of any local degrees of freedom in this theory. It thus becomes clear that asymptotically AdS${}_3$ spacetimes are parametrized by
 certain pairs of functions of $t\pm\theta$. More explicitly, the work of Brown and Henneaux \cite{Brown:1986nw} shows that the group of asymptotic symmetries of asymptotically
 AdS spacetimes is given by two copies of $\Diff_+(\bbS^1)$. The phase space of asymptotically AdS${}_3$ spacetimes could then be described as the quotient space of this group
 modulo the AdS${}_3$ isometry group, giving overall two copies of $\Diff_+(\bbS^1)/{\rm SL}(2,\R)$, see e.g. Sect.2.2 of \cite{Maloney:2007ud}. 

The above description is, however, not entirely satisfactory. Indeed, the Fefferman-Graham coordinate $\rho$ extends only over a portion of the spacetime near
 its conformal boundary. Thus, only very little control over what happens inside the spacetime is available. In particular, it is not possible to know whether
 a spacetime (\ref{intr-FH}) contains any non-trivial topology. It is also very hard to characterize those choices of $a_\pm$ that lead to non-singular spacetimes.
 For all these reasons the description (\ref{intr-FH}), while indicating that there is some infinite-dimensionality to be expected, is not a satisfactory description
 of the phase space of asymptotically AdS${}_3$ gravity.

As we will show in this paper, a description of the phase space that overcomes these difficulties is possible by using embedded maximal surfaces. Indeed, one
 particularly powerful description of the compact spatial slice situation is based on maximal surfaces, see \cite{Krasnov:2005dm}. It can be shown that each
 AdS${}_3$ spacetime with a compact spatial slice (such spacetimes were referred to as globally hyperbolic maximal compact (GHMC) AdS in \cite{Krasnov:2005dm})
 contains a unique maximal surface. The first and second fundamental forms induced on such a surface then become the configurational and momentum variables. It
 can be shown that the free data are those of a conformal structure and a certain quadratic differential on the maximal surface, and these together parametrize
 a point in the cotangent bundle of the Teichm\"uller space of the Cauchy surface. The data on the maximal surface can in turn be used to produce two hyperbolic
 metrics via a generalized Gauss map, and thus two points in the Teichm\"uller space, and this way one obtains an explicit realization of the Mess parametrization
 \cite{Mess:2007}. 

The present work more or less generalizes the above compact case description to the non-compact setting. Thus, similar to the construction described in
 \cite{Krasnov:2005dm}, we shall present two parametrizations of the phase space. One of them works with conformal structures and quadratic differentials on the disc,
 and thus provides an analogue of the cotangent bundle description. The other works with two conformal structures on the disc, and is the analogue of the two copies of
 the Teichm\"uller space description. The relation between both parametrizations is obtained explicitly from the harmonic decomposition of quasiconformal
 minimal Lagrangian diffeomorphisms of the disc, and generalizes the Mess to a symplectomorphism $T^\ast\T(1)\rightarrow\T(1)\times\T(1)$. Our main
 mathematical result is a description of this highly non-trivial map, and a proof of the fact that it is bijective. We also give a ``physicist's'' proof
 that this map is a symplectomorphism making use of the Chern-Simons description of 2+1 gravity.

We would like to emphasize that our description of the phase space of AdS${}_3$ gravity by two copies of the universal Teichm\"uller includes and in a sense supercedes the
 description by two copies of $\Diff_+(\bbS^1)/{\rm SL}(2,\R)$ that follows from the Brown-Henneaux work, see e.g. \cite{Maloney:2007ud}, Sect.2.2, for a particularly clear
 account of the Brown-Henneaux parameterization of the phase space. Indeed, as we shall see below, in one possible description $\T(1)$ is realized as the space
 $\QS(\bbS^1)/{\rm SL}(2,\R)$ of (normalized) quasisymmetric homeomorphism of $\bbS^1$, and this contains $\Diff_+(\bbS^1)/{\rm SL}(2,\R)$ as a submanifold. However,
 more important than the extension from the space of smooth to that of quasisymmetric maps, our construction also describes how the bulk moduli (non-trivial topology) can
 be encoded in the phase space.

The constructions in this paper builds on and extends those in \cite{Bonsante:2010} and \cite{Aiyama:2000}.
 Thus, there is not much new mathematics in this work. Rather, we take some results obtained by mathematicians (with different aims), and use them to describe
 the phase space of AdS${}_3$ gravity. In particular, our description is based on the result in \cite{Bonsante:2010} that proved the existence and uniqueness
 of maximal surfaces in AdS${}_3$ with a given boundary curve. This boundary curve is, in turn, parametrized by a single quasisymmetric homeomorphism on the
 circle, so there is a one-to-one correspondence between quasisymmetric homeomorphisms of the circle (i.e. points in the universal Teichm\"uller space) and
 maximal surfaces in AdS${}_3$. We use these, as well as some results from the universal Teichm\"uller literature, to describe the phase space, seen as the
 space of all AdS${}_3$ spacetimes, in terms of deformations of the domain of dependence of a totally geodesic spacelike surface in AdS${}_3$.

\begin{figure}
\label{fig1}
\begin{center}
\includegraphics[width=150mm]{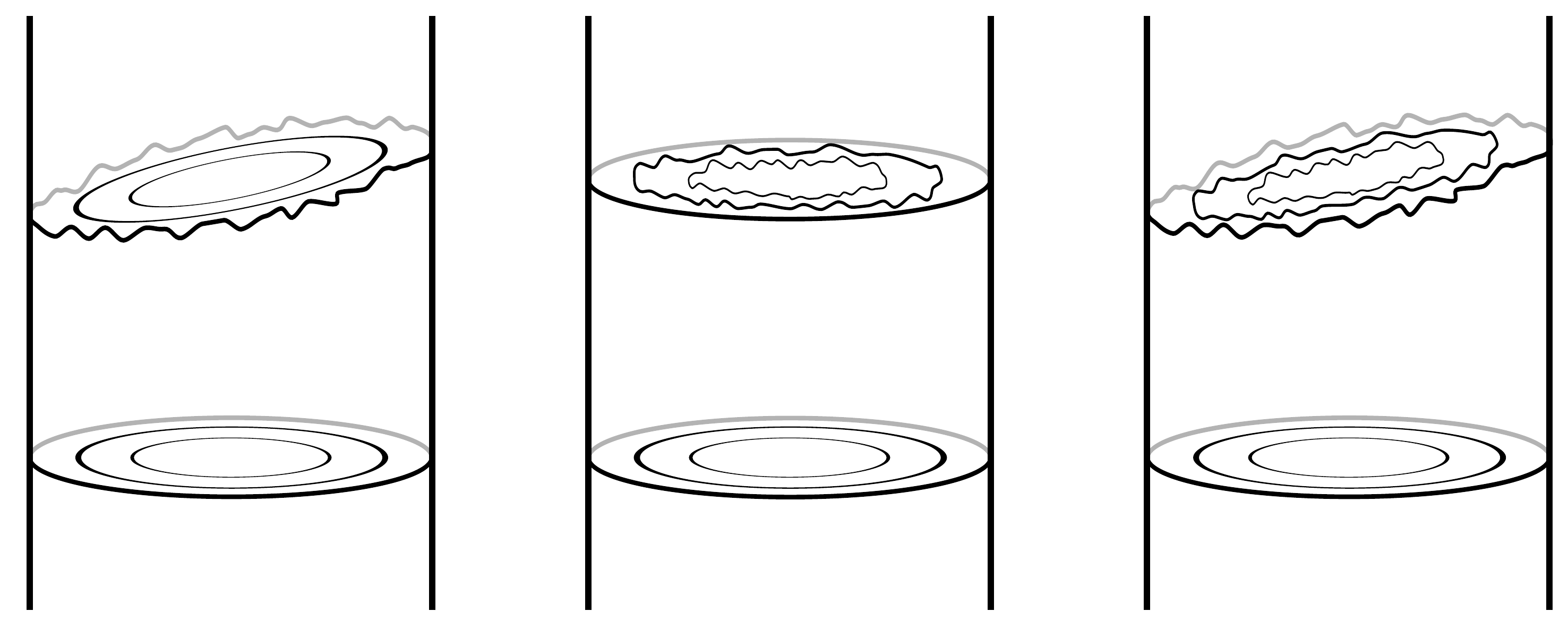}
\caption{Two deformation directions. One (left figure) corresponds to deforming the curve along which the maximal surface intersects the boundary. The other
 (middle figure) corresponds to deforming the complex structure on the maximal surface, or, geometrically, to deforming the constant ``radial'' coordinate
 foliations of the surface. A general point in the phase space deforms both the curve at infinity as well as the constant radial coordinate foliation of the
 maximal surface (right figure).}
\end{center}
\end{figure}

One non-obvious point of our construction, which is also where we depart from the works cited above, is the existence of two independent directions in the phase
 space. Thus, the work \cite{Bonsante:2010} makes it clear that a single point in $\T(1)$ gives rise to a maximal surface in AdS${}_3$, which then comes equipped with its
 first and second fundamental forms (induced by the embedding in AdS${}_3$). Thus, it can appear that a single quasisymmetric homeomorphism on the circle (single
 point in $\T(1)$) is sufficient to specify all of the ``initial'' data necessary for the maximal surface description. The question is then where does the second
 phase space direction, i.e. a second point in $\T(1)$, comes from. As we shall describe in more details below, this other direction comes from the possibility of
 an additional quasiconformal deformation on the maximal surface. Being a diffeomorphism it does not change the first and second fundamental forms on the maximal
 surface, but being asymptotically non-trivial it gives rise to deformations that has to be considered as non-gauge. And we shall verify that the two types of
 deformations --- the geometric ones corresponding to changing the curve along which the maximal surface intersects the boundary, and the non-geometric one
 corresponding to just performing an asymptotically non-trivial diffeomorphism on the maximal surface --- are canonically conjugate to each other in the symplectic
 structure induced by the gravitational action. Thus, both are equally important as far as AdS${}_3$ gravity is concerned. The two phase space directions --- those
 deforming the curve along which the maximal surface intersects the cylinder at infinity and those deforming the complex coordinate on the maximal surface --- are
 graphically depicted in Fig.\ref{fig1}.

Another, more geometrical description of our phase space was suggested to us by Jean-Marc Schlenker after the first version of this paper was put on
 the arxiv \cite{JM}. Thus, the phase space can be described as the space of quasiconformal maximal embeddings of the unit disc $\Delta$ into AdS${}_3$,
 considered up to a natural in this context identification: two quasi-conformal maps $u,v: \Delta \to {\rm AdS}_3$ are considered equivalent if their
 composition $u^{-1} \circ v$ is the identity at infinity. 

As in the compact setting, the parametrization by two copies of the (universal) Teichm\"uller space is related to the Chern-Simons formulation of 2+1 gravity
 introduced in \cite{Witten:1988hc}. Indeed, we recall that, for AdS spacetimes, the Einstein-Hilbert action can be written as two copies of ${\rm SL}(2,\R)$
 Chern-Simons action. Every AdS metrics can, therefore, be described by an associated pair of flat ${\rm SL}(2,\R)$ connections. Then a simple explicit computation
 shows that, in our parametrization, each copy of $\T(1)$ corresponds to one of these connections. In fact, this relation to the CS formulation is the easiest way
 to understand why the generalized Mess map  $T^\ast\T(1)\rightarrow\T(1)\times\T(1)$ is a symplectomorphism, see below for a further discussion of this point.

The outline of the present paper is as follows. In Sect.\ref{sec:compact} we give a brief review of the compact case. Section \ref{sec:maxsurf} deals with the
 maximal surfaces in AdS${}_3$. The construction of the phase space of globally hyperbolic AdS spacetimes is described in Sect.\ref{sec:mess}. We present the
 generalized Mess map in Sect.\ref{sec:map}. A relation to the holographic description is worked out in Sections \ref{sec:infinitesimal}, \ref{sec:bers},
 \ref{sec:charges}. We finish with a discussion. For those not familiar with (universal) Teichm\"uller theory we present a quick overview in the Appendix.

\section{Compact Spatial Topology}
\label{sec:compact}
In this Sect.we consider globally hyperbolic AdS${}_3$ spacetimes $M=\R\times\Sigma$ whose Cauchy surface is a genus $g\geq2$ Riemann surface $\Sigma$.
 In the Hamiltonian formulation of general relativity one starts by foliating spacetime by spacelike hypersurfaces. The spacetime metric is then described
 in terms of the first and second fundamental forms $(I,\II)$ of the initial Cauchy surface $\Sigma$. The first and second fundamental forms have to satisfy
 certain relations, known as the Gauss-Codazzi equations (or as the Hamiltonian and momentum constraints in the GR community). In \cite{Moncrief:1989dx}, the
 phase space of flat 2+1 gravity was shown to be parametrized by the cotangent bundle over Teichm\"uller space of the initial surface by choosing a foliation
 by constant mean curvature surfaces. In fact, in isothermal coordinates on $\Sigma$, associated with its conformal structure, we can write
$$I=e^{2\varphi}|dz|^2,\quad \II=\frac{1}{2}(qdz^2+\bar qd\bar z^2)+H e^{2\varphi}|dz|^2,$$
where $H$ is the mean curvature of $\Sigma$. Then, Codazzi equation imposes holomorphicity of the quadratic differential $qdz^2$ defined by the traceless
 part of $\II$ and Gauss equation becomes an equation for the ``Liouville'' field $\varphi$. This equation becomes particularly simple on a maximal surface
 $H=0$ and reads:
\be\label{Gauss}
4\partial_z\partial_{\bar z}\varphi=e^{2\varphi}-e^{-2\varphi}|q|^2.
\ee
Note that this is the equation relevant for the AdS${}_3$ setting. Similar equations (with some sign changes) hold also in the positive or zero scalar curvature
 settings. The important fact is that the existence and uniqueness of the solution of the above Gauss equation holds, implying the existence and uniqueness of a
 maximal surface in any globally hyperbolic AdS${}_3$ spacetime (here with compact spatial slices). Given this existence and uniqueness result, the pair $(I,\II)$
 is completely determined by a conformal structure $z$ and a holomorphic quadratic differential $qdz^2$, thus a point in $T^\ast\T(\Sigma)$. This gives an efficient
 explicit description of the spacetime geometry as parametrized by the data on the maximal surface. Indeed, the 3-metric can be written in the equidistant coordinates
 to the maximal surface as
\be\label{metric}
ds^2=-d\tau^2+\cos^2\tau I+2\sin\tau\cos\tau\II+\sin^2\tau\II\,I^{-1}\II.
\ee
A direct computation then shows the gravitational symplectic structure agrees with the canonical cotangent bundle one, see \cite{Krasnov:2005dm} for more details.
 We thus have a description of the phase space of AdS${}_3$ gravity in the compact spatial slice setting as $T^\ast \T(\Sigma)$.

In \cite{Mess:2007} Mess obtained another parametrization of the same spacetimes by two copies of Teichm\"uller space $\T(\Sigma)\times\T(\Sigma)$. His construction
 can be understood as follows. In the projective model, AdS${}_3$ can be seen as the image of the quadric $X=\{x\in\R^{2,2};\langle x,x\rangle=-1\}$, with its induced
 metric, under the projection $\pi:X\rightarrow\R P^3$, 
$${\rm AdS}_3=\pi(X)=\{[x]\in\R P^3;\langle x,x\rangle<0\}.$$
A point on AdS${}_3$ is then in correspondence with a line in $\R^{2,2}$ passing though a point in the quadric $X$ and the origin. The boundary of AdS${}_3$, the
 projective quadric $$\partial{\rm AdS}_3=\{[x]\in\R P^3;\langle x,x\rangle=0\},$$ is known to be foliated by two families of projective lines $\mathcal{L}_+$ and
 $\mathcal{L}_-$ (corresponding to the left and right null geodesics). Since each line of one family intersects a line of the other family a single time, this
 provides an identification of $\partial{\rm AdS}_3$ with the torus $\bbS^1\times\bbS^1$.

Now, a line in $\mathcal{L}_+$ (resp. $\mathcal{L}_-$) meets the boundary of any spacelike surface at a single point so one can use these ``left'' and ``right''
 families to define maps between the boundaries of any pair of spacelike surfaces. Let us first require these spacelike surfaces to be totally geodesic. Given a
 pair of such totally geodesic spacelike surfaces $P_0$ and $P$, let $\pi_+,\pi_-:\partial P\rightarrow \partial P_0$ be the ``left'' and ``right'' maps of their
 boundaries. These then uniquely extend to isometries $\Phi_+^P,\Phi_-^P:P\rightarrow P_0$ of AdS${}_3$ sending $P$ to $P_0$. Now, taking an arbitrary spacelike
 surface $S$, one can associate to $S$ a pair of diffeomorphisms $\Phi_+,\Phi_-:S\rightarrow P_0$ by
$$\Phi_+(x)=\Phi_+^{P(x)}(x),\quad\Phi_-(x)=\Phi_-^{P(x)}(x),$$
where $P(x)$ is the totally geodesic spacelike surface tangent to $S$ at $x$. Taken modulo an overall isometry in AdS${}_3$, this construction is independent of
 the choice of $P_0$.

Given an GHMC AdS spacetime $M$, let $\Sigma$ be some smooth embedded spacelike surface and consider its lift $S$ into the universal cover of $M$, which is AdS${}_3$.
 Taking the pull-back of the hyperbolic metric on $P_0$ by $\Phi_+$ and $\Phi_-$ defines two hyperbolic metrics on $S$, which in turn descend to hyperbolic metrics
 $I_+$ and $I_-$ on $\Sigma$, thus defining a point in $\T(\Sigma)\times\T(\Sigma)$. This construction can be applied to any spacelike surface $\Sigma$ in $M$ and it
 can be shown that the point in $\T(\Sigma)\times\T(\Sigma)$ one gets is independent of this choice. When $\Sigma$ is maximal (or at least of constant mean curvature),
 the data on $\Sigma$ also gives rise to a point in $T^\ast \T(\Sigma)$, as we reviewed above, so this defines what can be referred to as the Mess map,
\be\label{mess}
{\rm Mess}:T^\ast\T(\Sigma)\rightarrow\T(\Sigma)\times\T(\Sigma)
\ee
taking $(I,\II)$ on the maximal surface into $(I_+,I_-)$. It can be shown that this map is a bijection, see \cite{Krasnov:2005dm}, so there are two equivalent
 descriptions of the moduli space of globally hyperbolic AdS${}_3$ spacetimes with spatial sections of fixed topology. One is given by $T^\ast \T(\Sigma)$, the
 other by $\T(\Sigma)\times \T(\Sigma)$.

By a calculation, also available in \cite{Krasnov:2005dm}, the Mess map can be explicitly described as follows 
\be\nonumber
I_\pm=I(E\pm JI^{-1}\II\,\cdot\,,\,E\pm JI^{-1}\II\,\cdot\,),
\ee 
where $E$ is the identity operator on $T\Sigma$ and $J$ is the complex structure associated with the conformal structure of $\Sigma$. These two metrics on $\Sigma$
 are hyperbolic provided $I,\II$ satisfy the Gauss and Codazzi equations. We note that each of the maps $\Phi_\pm$ is harmonic (with their Hopf differentials adding
 to zero) so the maps $\Phi_\pm$ can be referred to as the generalized Gauss map, see e.g. \cite{Aiyama:2000}. This terminology is legitimate given the resemblance
 of the construction of the metrics $I_\pm$ with the famous Gauss map between the data on a constant mean curvature surface in $\R^{2,1}$ and hyperbolic metrics.

The work \cite{Krasnov:2005dm} also described an explicit inverse of the Mess map, by providing a map between a pair $I_\pm$ of hyperbolic metrics on $\Sigma$ and
 the first and second fundamental forms of the {\it maximal} surface in the spacetime $M$ that corresponds to $I_\pm$. This map uses the existence and uniqueness of
 a minimal Lagrangian diffeomorphism between a surface $\Sigma$ equipped with the ``left'' and ``right'' hyperbolic metrics $I_\pm$. We shall denote this map by
 $F:(\Sigma,I_+)\rightarrow(\Sigma,I_-)$. It is an area preserving diffeomorphism (hence the term Lagrangian) whose graph is minimal in the product
 $(\Sigma\times\Sigma,I_+\times I_-)$ (hence the term minimal). As is reviewed in \cite{Krasnov:2005dm}, the existence of a minimal Lagrangian $F$ is equivalent
 to the existence of an operator $b:T\Sigma\rightarrow T\Sigma$ satisfying
\begin{enumerate}
 \item{$\det b=1$;}
 \item{$b$ is self-adjoint with respect to $I_+$;}
 \item{$d^{\nabla^+}b=0$, where $\nabla^+$ is the Levi-Civita connection of $I_+$;}
 \item{$F^\ast I_-=I_+(b\,\cdot\,,b\,\cdot\,)$.}
\end{enumerate}
In terms of $b$ one can construct a metric and a symmetric bilinear form on $\Sigma$ 
\be\label{inverse-mess}
I=\frac{1}{4}I_+(E+b\,\cdot\,,E+b\,\cdot\,),\qquad\II=-IJ(E+b)^{-1}(E-b)
\ee 
which satisfy the Gauss-Codazzi equations. Thus, the problem of constructing the inverse map reduces to the problem of determining the operator $b$. Once this is
 known, the first and second fundamental form $I,\II$ obtained by the above formulas are those of the {\it maximal} surface in the spacetime corresponding to the
 pair $I_\pm$. An explicit expression for the spacetime metric is then given by (\ref{metric}) providing an efficient parametrization of the space of globally
 hyperbolic AdS${}_3$ spacetime (with compact spatial slices) by two copies of the Teichm\"uller space. This description of the inverse of the Mess map admits a
 direct generalization to the non-compact setting of interest to us, and will play an important role in the next Sections. 

We also note that the gravitational symplectic structure, evaluated in the parametrization $T^\ast\T(\Sigma)$, is just the canonical cotangent bundle symplectic
 structure, see \cite{Krasnov:2005dm} for a simple calculation that demonstrates this. It can also be verified that the map ${\rm Mess}$ is a symplectomorphism with
 respect to the natural symplectic structure on $\T(\Sigma)\times\T(\Sigma)$ given by two copies of the Weil-Petersson symplectic structure. The easiest way to see
 this is to use the Chern-Simons description in which the left and right metrics of the Mess parametrization encode the monodromies of the left and right
 ${\rm SL}(2,\R)$ connections on $\Sigma$. Since the CS formulation provides an equivalent description of AdS${}_3$ gravity, and the symplectic structure
 of ${\rm SL}(2,\R)$ CS theory reduces to the Weil-Petersson symplectic structure on $\T(\Sigma)$, the Mess map must be symplectic. We shall see this explicitly
 (at the origin of both spaces) in Sect.(\ref{sec:sympl}).

\section{Maximal Surfaces in AdS${}_3$ and Universal Teichm\"uller Space}
\label{sec:maxsurf}
As a preparation for our consideration of the non-compact setting, we start by reviewing the relation between the universal Teichm\"uller space and maximal
 surfaces in AdS${}_3$ described in \cite{Bonsante:2010}. We shall also present some known facts relating maximal surfaces in AdS${}_3$, harmonic maps and
 minimal Lagrangian maps between the unit disc, see \cite{Aiyama:2000}. 

We start by reviewing some details about the universal Teichm\"uller space. We follow most conventions of \cite{Takhtajan:2006}, and the reader is advised to
 consult these references for more details. In general terms, the universal Teichm\"uller space is the quasiconformal deformation space of the unit disc
 $\Delta$ and can be realized as the space of certain equivalence classes of bounded Beltrami coefficients on $\Delta$ by solving Beltrami equation. More
 concretely, let
$$L^\infty(\Delta)_1=\left\{\mu:\Delta\rightarrow\C;|\mu|_\infty=\sup_{\Delta}|\mu(w)|<1\right\},$$
be the unit ball in the space of bounded Beltrami coefficients in the unit disc $\Delta$. Given $\mu,\nu\in L^\infty(\Delta)_1$ one solves the Beltrami equation
$$\tilde\mu = \partial_{\bar{w}} z/\partial_w z$$
in $\hat\C$ with coefficients extended by reflection
\be\nonumber \nonumber
\tilde\mu(w)=\begin{cases}\overline{\mu(1/\bar w)}w^2/\bar w^2,&w\in\hat\C\backslash\Delta,\cr
\mu(w),& w\in\Delta\end{cases}
\ee
similarly for $\nu$. The equivalence relation between $\mu,\nu$ is then given if the corresponding solutions, normalized to fix $-1$, $-\im$ and $1$, agree on
 the conformal boundary $\bbS^1$
\be\label{equiv}
z_\mu|_{\bbS^1}=z_\nu|_{\bbS^1}.
\ee
Since the boundary values of quasiconformal diffeomorphisms of the disc are quasisymmetric homeomorphisms of the unit circle, we have an identification between
 the universal Teichm\"uller space $\T(1)$ and the space ${\rm QS}(\bbS^1)/{\rm SL}(2,\R)$ of M\"obious normalized quasisymmetric homeomorphisms of $\bbS^1$. For
 purposes of this Section it will be sufficient to think about $\T(1)$ as such space.

This is the so-called A-model of the universal Teichm\"uller space. The complex structure in $\T(1)$ is not easily described in this model, but become much more
 explicit in another description, the so-called B-model, which works with solutions of the Beltrami equation where the Beltrami coefficient is extended to the
 outside of the unit disc as $\mu=0$. Some more facts about the two models and their relation are described in Section \ref{sec:bers}.

We note that the universal Teichm\"uller space has a (formal) symplectic manifold structure generalizing the Weil-Petersson symplectic structure of the closed
 topology Teichm\"uller spaces. The symplectic structure is described in some detail in the Appendix, but we refer the reader to a more complete exposition in
 \cite{Takhtajan:2006}. 

Before we explain a relation between $\T(1)$ and the maximal surfaces in AdS${}_3$, we would like to describe how the fixed topology Teichm\"uller spaces $\T(\Sigma)$
 can be realized as submanifolds in $\T(1)$, thus justifying the nomenclature used in this theory. This is achieved by restricting the Beltrami coefficients introduced
 above to have fixed periodicity properties with respect to some ``base point'' Fuchsian group. Thus, let $\Gamma$ be a fixed Fuchsian group such that the quotient
 $\Delta/\Gamma$ is a Riemann surface of the required topology. We then define a space of Beltrami coefficients for $\Gamma$
\be\nonumber
L^\infty(\Delta,\Gamma)_1=\left\{ \mu\in L^\infty(\Delta)_1: \mu\circ \gamma \frac{\overline{\gamma'}}{\gamma'} = \mu, \forall \gamma\in \Gamma\right\}
\ee
and the corresponding Teichm\"uller space of $\Sigma=\Delta/\Gamma$ is obtained as
\be\nonumber
\T(\Sigma)=\T(\Gamma)=L^\infty(\Delta,\Gamma)_1/\sim,
\ee
where the equivalence relation is the same as the one introduced above, see (\ref{equiv}). It can be shown that this is just the space of all fixed topology
 Fuchsian groups, which arise as
\be\nonumber
\Gamma_\mu = z_\mu \circ \Gamma \circ z_\mu^{-1}.
\ee
Thus, we have given a description of the Teichm\"uller space $\T(\Sigma)$ as the quasiconformal deformation space of $\Gamma$. This can be seen as a ball of radius
 one in the space of ($\Gamma$-periodic) Beltrami coefficients centred at the preferred surface $\Delta/\Gamma$. Note that the group $\Gamma$ can be chosen to be
 rather arbitrary here. One possible choice is that $\Delta/\Gamma$ is a compact surface of given genus. However, a choice where $\Delta/\Gamma$ is an infinite area
 surface with hyperbolic ends is also possible. This latter choice is the one relevant for the description of multi-black-holes.

We would also like to note, without going into much details, that it is possible to generalize the discussion to  include the case of non-orientable spatial topology.
 Fuchsian groups should then be replaced by the so-called non-euclidean crystallographic (NEC) groups, discrete groups of isometries of $\Delta$ including orientation
 reversing elements. The invariance property of Beltrami coefficients under these additional elements should then be $\mu\circ\gamma (\overline{\gamma'}/\gamma') =\bar\mu$.
 Then every Klein  surface $\Sigma$ has an orientable complex double cover $\Sigma^c$ and the Teichm\"uller space of $\Sigma$ embeds as an open submanifold of
 $\T(\Sigma^c)$. We refer the reader to \cite{Seppala:1992} for an exposition on Kleinian surfaces. Thus, if desired, the universal construction of this paper also
 includes the ``geon'' spacetimes studied in e.g. \cite{gr-qc/9906031}.

We now describe a relation between points in $\T(1)$ and maximal surfaces in AdS${}_3$, established in \cite{Bonsante:2010}. The key idea here is that, given a
 quasisymmetric map $\bbS^1\to \bbS^1$, its graph in $\bbS^1\times\bbS^1$ can be viewed as a spacelike curve on the conformal boundary $\partial {\rm AdS}_3$.
 Indeed, as we recalled in the previous Section, $\partial{\rm AdS}_3$ is ruled by two families of left and right null geodesics, and is therefore (a 2-to-1 cover of)
 $\bbS^1\times \bbS^1$. Now, given a spacelike curve on $\partial{\rm AdS}_3$, it is shown there is a unique maximal surface in AdS${}_3$ intersecting the conformal
 boundary along this curve. The existence part here is quite general and makes only very weak assumptions about the curve at infinity. It is in the proof of uniqueness
 that quasisymmetric boundary curves become relevant. Thus, \cite{Bonsante:2010} introduces the notion of a width $w(\Gamma)$ of the boundary curve $\Gamma$ which is
 the supremum of the (time) distance between the upper and lower boundaries of its convex hull convex hull $C(\Gamma)$ in AdS${}_3$. In one hand, this provides a
 new characterization of quasisymmetric maps in terms of AdS${}_3$ geometry. It is shown that for any boundary curve the width is at most $\pi/2$, being strictly
 less than $\pi/2$ if and only if it is the graph of a quasisymmetric map $\bbS^1\to \bbS^1$. On the other hand it gives sufficient conditions for the uniqueness
 result. First, the condition $w(\Gamma)<\pi/2$ is shown to imply that the corresponding maximal surface has sectional curvature bounded above by a negative
 constant. Then, based on convexity properties, it is shown that this maximal surface with uniformly negative sectional curvature is unique among complete
 maximal graphs with the given boundary curve and bounded sectional curvature.

The existence and uniqueness of maximal surfaces in AdS${}_3$ corresponding to points in $\T(1)$, i.e. quasisymmetric maps, can then be seen equivalent to the
 existence and uniqueness of quasiconformal minimal Lagrangian extensions of quasisymmetric homeomorphisms of $\bbS^1$ to the interior of the disc. This last
 point is essentially the same construction as occurs in the compact setting, see the previous Section, where a maximal surface in AdS${}_3$ gives rise to a
 minimal Lagrangian diffeomorphism between the Riemann surfaces $(\Sigma, I_+)$ and $(\Sigma, I_-)$. The same construction now extends to the non-compact
 setting, as was used in \cite{Aiyama:2000}, and allowed \cite{Bonsante:2010} to prove the existence and uniqueness of minimal Lagrangian extensions of
 quasisymmetric maps.

\section{The Generalized Mess Parametrization of the AdS${}_3$ Phase Space}
\label{sec:mess}
In this Section we describe the universal phase space of AdS spacetimes as parametrized by two copies of the universal Teichm\"uller space. This is
 essentially a generalization of the inverse of the Mess map, as described towards the end of Sect.\ref{sec:compact}. The moduli space of fixed
 spatial topology AdS${}_3$ manifolds is then obtainable from the universal phase space by restricting the Beltrami coefficients to be invariant under
 appropriate topology discrete subgroups of ${\rm SL}(2,\R)$. In Sect.\ref{sec:infinitesimal} we shall see how the ``non-geometrical'' asymptotic
 degrees of freedom of Brown-Henneaux are encoded in this phase space.

Thus, let us take two points in $\T(1)$, which, we remind the reader, can be thought of as two (normalized to fix $-1,1,-\im$) quasisymmetric homeomorphisms
 of the circle. Let us denote these homeomorphisms by $f_\pm: \bbS^1\to \bbS^1$. As will become clear below, it will be useful to think about $f_\pm$ as the
 boundary values of two quasiconformal maps $z_\pm$ from a ``base point'' unit disc representing a preferred base point in $\T(1)$. We call the complex
 coordinate on the base disc $w$, see Fig.2, and interpret the maps $z_\pm$ as deformations of this disc defining complex coordinates, which we also call
 $z_\pm$, on the target discs $\Delta_\pm$. Each of the discs $\Delta_\pm$ has its standard hyperbolic metric which we denote
\be\nonumber
I_\pm=\frac{4|dz_\pm|^2}{(1-|z_\pm|^2)^2}.
\ee
Note that $I_\pm$, although both hyperbolic, are to be considered as representatives of inequivalent conformal structures of the disc. Although $I_\pm$ are in
 fact isometric, the isometry mapping $I_+$ to $I_-$ acts nontrivially at infinity changing the equivalence class. The choice of a base point unit disc thus
 becomes quite helpful in avoiding confusion. In fact, when pulled back to the base disc the metrics $I_\pm$ explicitly involve the Beltrami coefficients
 associated with points in $\T(1)$ they represent
$$I_\pm=\frac{4|\partial_wz_\pm|^2}{(1-|z_\pm(w)|^2)^2}|dw+\mu_\pm d\bar w|^2,$$
where $\mu_\pm=\partial_{\bar{w}} z_\pm /\partial_w z_\pm$. It is in this sense that we use the hyperbolic metrics $I_\pm$ on $\Delta$ as representatives of
 points in $\T(1)$.

A word is in order about which quasiconformal extensions $z_\pm$ of quasisymmetric boundary maps $f_\pm$ are considered. Indeed, there are many quasiconformal
 maps $z_\pm$ in the same equivalence class in the sense of universal Teichm\"uller theory, i.e. having the same restrictions $f_\pm$ to the circle. We shall
 see below that for our purposes the composition $z_-\circ z_+^{-1}$ will need to satisfy certain property which makes it unique, given the boundary values.
 Apart from this restriction, the extensions $z_\pm$ are arbitrary, and for our construction of the phase space it will not matter which specific extension is
 chosen. 

\begin{figure}
\label{fig}
\begin{center}
\includegraphics[width=80mm]{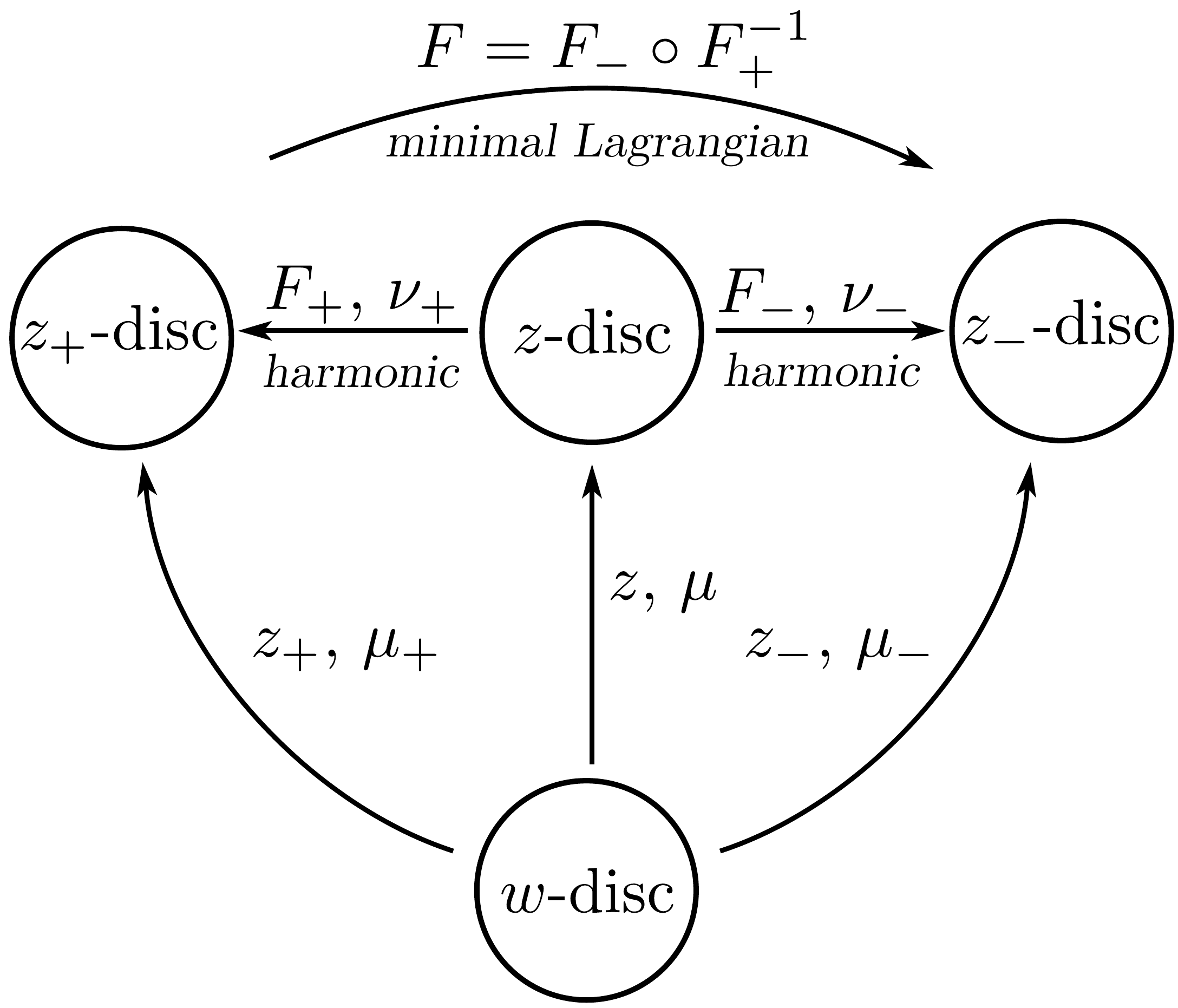}
\caption{The diagram of maps}
\end{center}
\end{figure}

Let us now consider $f=f_-\circ f_+^{-1}$ obtained by composing the homeomorphisms $f_\pm$. This is also a quasisymmetric homeomorphism of $\bbS^1$, and
 according to \cite{Bonsante:2010} there is a unique  maximal surface (with uniformly negative sectional curvature) in AdS${}_3$ whose boundary curve is
 the graph of $f$. Let $F:\Delta_+\to \Delta_-$ denote the minimal Lagrangian extension of $f$ obtained in \cite{Bonsante:2010}. We now fix the arbitrariness
 in $z_\pm$ (to some extent) by requiring their composition to agree with this minimal Lagrangian diffeomorphism
\be\nonumber
F= z_-\circ z_+^{-1}.
\ee
Note that this is a condition on the maps $z_\pm$ rather than on $F$, the latter being completely fixed by the boundary quasisymmetric maps $f_\pm$.

As in the compact case, the knowledge of $F$ and the hyperbolic metric on e.g. the disc $\Delta_+$ are enough to reconstruct both the first and second
 fundamental forms on the maximal surface. Formula (\ref{inverse-mess}) gives their expressions evaluated at the $z_+$-disc. This description, however,
 completely hides one direction in the phase space. To obtain the complete description with the dependence on both $(f_+,f_-)$ explicit, we take the
 pull-back to the base disc. The nature of the two deformation directions in the phase space then becomes clear. Their ``difference'' $f=f_-\circ f_+^{-1}$
 defines a maximal surface in AdS${}_3$, with its induced first and second fundamental forms and the corresponding isothermal coordinate. We interpret this
 as a ``geometric'' direction as it completely describes the geometry of the maximal surface, and of its Cauchy development. The other direction, corresponding
 to the particular way $f$ decomposes into $f_\pm$, determines another quasiconformal deformation of the maximal surface (or a choice of the complex coordinate
 on the maximal surface possibly different from the isothermal coordinate $z$ there). This measures how distant our spacetime is from the preferred base point
 in the phase space and can interpreted as a ``non-geometric'' deformation determining a constant radius foliation of the maximal surface.

To write down the first and second fundamental forms on the maximal surface in terms of the complex coordinate measuring how distant our spacetime $w$ on the base
 disc, we need to describe the operator $b:T\Delta_+\rightarrow T\Delta_+$ associated with the minimal Lagrangian diffeomorphism $F$. Then, the inverse Mess map
 (\ref{inverse-mess}) directly gives the first and second fundamental forms of the corresponding maximal surface. It is possible to write $b$ explicitly as 
$$b=|\partial F|\left(\partial_{z_+}dz_+ +\lambda\partial_{z_+}d\bar{z}_+ +\bar\lambda \partial_{\bar{z}_+}dz_+ +\partial_{\bar{z}_+}d\bar{z}_+\right)$$
where $\lambda=\partial_{\bar z_+}F/\partial_{z_+}F$ is the Beltrami coefficient of $F$ and
$$|\partial F|=\frac{1-|z_+|^2}{1-|F|^2} |\partial_{z_+} F|$$ 
is the holomorphic energy density of $F$. When $F$ is minimal Lagrangian this operator satisfies the conditions (1-4) in the definition of the inverse of the
 Mess map and we can construct the fundamental forms of the maximal surface as in the compact case
$$I=\frac{|\partial F|(|\partial F|+1)}{(1-|z_+|^2)^2}\left(2|dz_+|^2+\bar\lambda dz_+^2+\lambda d\bar z_+^2\right).$$
One then computes the almost-complex structure of $I$ and the operator $(E+b)^{-1}(E-b)$ directly $$J=\im|\partial F|\left(\partial_{z_+}dz_+ +\lambda\partial_{z_+}d\bar{z}_+-\bar\lambda \partial_{\bar{z}_+}dz_+-\partial_{\bar{z}_+}d\bar{z}_+\right)$$ and
$$(E+b)^{-1}(E-b)=-\frac{|\partial F|}{|\partial F|+1}\left(\lambda\partial_{z_+}d\bar{z}_++\bar\lambda \partial_{\bar{z}_+}dz_+\right).$$
Therefore the second fundamental form $\II$ is given by
$$\II=-\im\frac{|\partial F|}{(1-|z_+|^2)^2}(\bar\lambda dz_+^2-\lambda d\bar z_+^2).$$
These determine the first and second fundamental forms on the maximal surface, written in terms of the coordinate $z_+$ on the disc $\Delta_+$.

We now pull-back $(I,\II)$ to the base $w$-disc using the map $z_+$. The transformation of the Beltrami coefficient $\lambda$ and the holomorphic energy density
 $|\partial F|$ are easily obtained computing derivatives of $z_-=F\circ z_+$
$$\lambda\circ z_+\frac{\partial_{\bar w}\bar z_+}{\partial_wz_+}=\frac{\mu_--\mu_+}{1-\mu_-\bar\mu_+},\qquad |\partial F|\circ z_+|\partial_w z_+|=|\partial z_-|\frac{|1-\bar\mu_+\mu_-|}{1-|\mu_+|^2}.$$
Using the area preserving condition for $F$, we then get the following expressions
\begin{align}\label{I}
I=&\;\frac{1}{(1-|w|^2)^2}\left(\frac{1}{2}\frac{|\partial z_+|^2}{1-|\mu_-|^2}+\frac{1}{2}\frac{|\partial z_-|^2}{1-|\mu_+|^2}+\frac{|\partial z_+||\partial z_-|}{|1-\mu_-\bar\mu_+|}\right)\Big(2(1-|\mu_-|^2|\mu_+|^2)|dw|^2
\cr
&+(\bar\mu_+(1-|\mu_-|^2)+\bar\mu_-(1-|\mu_+|^2))dw^2+(\mu_+(1-|\mu_-|^2)+\mu_-(1-|\mu_+|^2))d\bar w^2\Big)
\end{align}
and
\begin{align}\label{II}
\II=&\;\frac{\im}{(1-|w|^2)^2}\frac{|\partial z_+||\partial z_-|}{|1-\mu_-\bar\mu_+|}\Big(2(\mu_+\bar\mu_--\bar\mu_+\mu_-)|dw|^2
+(\bar\mu_+(1+|\mu_-|^2)
\cr
&-\bar\mu_-(1+|\mu_+|^2))dw^2-(\mu_+(1+|\mu_-|^2)-\mu_-(1+|\mu_+|^2))d\bar w^2\Big).
\end{align}
Here 
\be\nonumber
|\partial z_\pm|=\frac{1-|w|^2}{1-|z_\pm|^2}|\partial_w z_\pm|, 
\ee
are the holomorphic energy densities of $z_\pm$. The pair $(I,\II)$ satisfies the Gauss-Codazzi equations, and can be used to construct the spacetime
 metric via (\ref{metric}). 

We close our discussion of the parametrization of AdS${}_3$ spacetimes by two copies of $\T(1)$ with a more detailed discussion of the ambiguity that entered
 into the above construction. Indeed, recall that the maps $z_\pm$ are arbitrary quasiconformal extensions of the boundary quasisymmetric maps $f_\pm$, with
 the condition that $z_-\circ z_+^{-1}$ is the minimal Lagrangian diffeomorphism extending $f_-\circ f_+^{-1}$. One can now see that nothing depends on the
 remaining extension ambiguity. Indeed, choosing two different extensions for $f_\pm$, say $\tilde z_\pm$, which are in the same universal Teichm\"uller class
 as $z_\pm$ and still satisfy $\tilde z_-\circ\tilde z_+^{-1}=F$, we obtain two pairs $(I,\II)$ and $(\tilde I,\tilde\II)$, as well as the corresponding
 spacetime metrics. It is however clear that the corresponding spacetimes can be mapped into one another by the (purely spatial) diffeomorphism $\tilde z_+\circ z_+^{-1}$,
 which is clearly asymptotically trivial since their restriction to the boundary is given by $$\tilde z_+\circ z_+^{-1}=f_+\circ f_+^{-1}=\Id.$$ Therefore these
 spacetimes should be considered equivalent and nothing in the above construction depends on which particular extension of $f_\pm$ are chosen, provided the minimal
 Lagrangian condition on $z_-\circ z_+^{-1}$ holds. 

\section{The Generalized Mess Map}
\label{sec:map}

The aim of this Section is to describe an analogue of the cotangent bundle parametrization of the universal phase space. To this end, we first verify the existence of a special
 decomposition of minimal Lagrangian diffeomorphisms in terms of harmonic maps with opposite Hopf differentials. The corresponding coordinate on the source disc of these maps is
 then shown to agree with the isothermal coordinate on the maximal surface and the Hopf differentials are just ($\im$ times) plus or minus the quadratic differential parametrizing
 the cotangent direction. These facts are not new and can be found in \cite{Aiyama:2000}. We give them here for completeness. We then describe what can be called a generalized
 Mess map from $T^\ast \T(1)$ to two copies of $\T(1)$. This arises precisely in the same way as in the compact setting, see (\ref{mess}), the only non-trivial point being the
 existence and uniqueness of a solution to the Gauss equation (\ref{Gauss}), which follows from results of \cite{Wan:1992}. In Section \ref{sec:sympl} we give a physicist's proof
 that this map is a (formal) symplectomorphism with respect to the natural symplectic structures on $T^\ast\T(1)$ and $\T(1)\times\T(1)$. The arising description of a
 symplectomorphism between these spaces is a new result, as far as we are aware. 

Any minimal Lagrangian diffeomorphism $F$ of the unit disc can be uniquely written as a composition $F=F_-\circ F_+^{-1}$ of two harmonic maps $F_\pm$ whose Hopf differentials
 add up to zero, see e.g. Lemma 2.1 of \cite{Aiyama:2000}. As we shall now see, it is this decomposition that leads to the particularly simple expressions for the data on the
 maximal surface. Let us refer to the coordinate on the source disc of $F_\pm$ by $z$, see Fig.2. The Hopf differentials are then given by
\be\nonumber
{\rm Hopf}(F_\pm)=\frac{4\partial_zF_\pm\partial_z\bar F_\pm}{(1-|F_\pm|^2)^2} dz^2.
\ee
Using the associated Beltrami differentials
$$ \nu_\pm = \partial_{\bar{z}} F_\pm /\partial_z F_\pm,$$
we can write these Hopf differentials as
\be\nonumber
{\rm Hopf}(F_\pm)= \frac{4|\partial F_\pm|^2\bar\nu_\pm}{(1-|z|^2)^2} dz^2,
\ee
where
$$|\partial F_\pm|=\frac{1-|z|^2}{1-|F_\pm|^2}|\partial_z F_\pm|$$
are the corresponding holomorphic energy densities. The Hopf differentials are required to add up to zero, which gives
$$\nu_+=-\frac{|\partial F_-|^2}{|\partial F_+|^2}\nu_-.$$
Then the area preserving condition for $F$ then reduces to
$$\frac{|\partial F_-|^4}{|\partial F_+|^4}|\nu_-|^2+\frac{|\partial F_-|^2}{|\partial F_+|^2}(1-|\nu_-|^2)-1=0,$$
and this implies
$$\frac{|\partial F_-|^2}{|\partial F_+|^2}=1,$$
in particular, $\nu_+=-\nu_-=\nu$. The fundamental forms (\ref{I}), (\ref{II}) of the maximal surface thus become
\be\label{I-II-max}
I=\frac{4|\partial F_+|^2}{(1-|z|^2)^2}|dz|^2,\qquad\II=\frac{\im}{2}\left({\rm Hopf}(F_+)-\overline{{\rm Hopf}(F_+)}\right).
\ee
Note that the functions $F_\pm$ are not holomorphic or anti-holomorphic in $z$, and so the metric $I$ is not hyperbolic, despite its seeming resemblance to
 $4|dF_+|^2/(1-|F_+|^2)^2$. The formulas (\ref{I-II-max}) are also contained in Proposition 3.1 of \cite{Aiyama:2000}. 

It is clear that what we have obtained is just the cotangent bundle description with the conformal factor and the holomorphic quadratic differential given by
$$e^{2\varphi}=\frac{4|\partial F_+|^2}{(1-|z|^2)^2},\qquad qdz^2=\im{\rm Hopf}(F_+).$$
We note that the Gauss-Codazzi equations for $\varphi$ and $q$ follow directly from the harmonicity of $F_+$
$$\partial_z\partial_{\bar z}F_++\frac{2 \overline{F_+}}{1- |F_+|^2}\partial_zF_+\partial_{\bar z}F_+=0.$$

We have thus seen that the Gauss maps composing the generalized Mess map (\ref{mess}), from the data on the maximal surface to the hyperbolic discs $\Delta_\pm$, are harmonic.
 This allowed for a simple description (\ref{I-II-max}) of the map $\T(1)\times \T(1)\to T^\ast \T(1)$. This fact also allows for a description of the map in the opposite
 direction. Thus, given a point $(\mu,qdz^2)\in T^\ast\T(1)$ one can solve for two harmonic maps $F_\pm$ with prescribed Hopf differentials
$$\im{\rm Hopf}(F_\pm)=\pm q.$$
 The existence and uniqueness of harmonic maps with prescribed Hopf differentials was given in \cite{Wan:1992} by proving that there exists a unique solution of the Gauss
 equation (\ref{Gauss}) such that the right- and left-hand-sides are non-negative and such that $e^{2\varphi} |dz|^2$ is a complete metric. This then allows to construct
 the harmonic maps explicitly via the Mess map (\ref{mess}). For more details of this construction the reader can consult \cite{Wan:1992}. We note that, although the treatment
 in this reference is carried for CMC surfaces in the Minkowski space $\R^{2,1}$, it needs very little adaptation to the present situation.

Finally, writing $\nu_\pm=\pm\nu$ for the corresponding Beltrami coefficients, it is now just a matter of using the group structure of $\T(1)$ to get the Beltrami coefficients
 of the maps from the base disc:
$$\mu_\pm=\frac{\mu\pm\nu\circ z (\partial_{\bar w}\bar z/\partial_wz)}{1\pm\bar\mu\nu\circ z(\partial_{\bar w}\bar z/\partial_wz)},\qquad \nu=\im\frac{(1-|z|^2)^2}{4|\partial F_\pm|^2}\bar q$$ This gives an explicit description of the generalized Mess map $T^\ast \T(1)\to \T(1)\times \T(1)$ and a proof it is a bijection.

\section{Relation to the Fefferman-Graham Description: The Infinitesimal Case}
\label{sec:infinitesimal}

In this Section we relate the description (\ref{metric}) of spacetimes as evolving data (\ref{I}), (\ref{II}) to the Fefferman-Graham type description of asymptotically AdS${}_3$
 spacetimes presented in the Introduction. Here we treat the infinitesimal case only, relating the objects appearing in our parametrization to the chiral functions $a_\pm$
 determining the Fefferman-Graham metric (\ref{intr-FH}). We shall accomplish this by computing the generators of both quasiconformal and asymptotic deformations of AdS$_3$,
 which we take to be the reference spacetime in both descriptions.

We start with the infinitesimal version of the spacetime metric in terms of initial data on the maximal surface. This is obtained by taking infinitesimal Beltrami coefficients
 $\delta\mu_\pm$ as representatives of the pair of quasisymmetric homeomorphisms $f_\pm\in\T(1)$. This can be interpreted in terms of infinitesimal quasiconformal deformations
 of the preferred base disc in $\T(1)$ in the direction of $\delta\mu_\pm$, now thought of as tangent vectors to $\T(1)$ at the origin. These deformations can be written
\be\nonumber
z_{\pm}=w+\delta z_\pm +\ldots,
\ee
where the first variations $\delta z_\pm$ are solutions of the first variations of Beltrami equation
\be\label{inf-beltrami}
\partial_{\bar w}\delta z_\pm=\delta\mu_\pm.
\ee
We refer the reader to the Appendix for more details on the tangent space to $\T(1)$.

It is then easy to obtain the corresponding infinitesimal versions of the data (\ref{I}) and (\ref{II})
\be\nonumber
I=\frac{4|dw|^2}{(1-|w|^2)^2}+\frac{2}{(1-|w|^2)^2}\Big[(\delta\bar\mu_++\delta\bar\mu_-)dw^2+(\delta\mu_++\delta\mu_-)d\bar w^2\Big]
\ee
\be\nonumber
\II=\frac{\im}{(1-|w|^2)^2}\Big[(\delta\bar\mu_+-\delta\bar\mu_-)dw^2-(\delta\mu_+-\delta\mu_-)d\bar w^2\Big]
\ee
and the infinitesimal metric
\begin{align}\label{metric-inf}
ds^2=&\;ds^2_{{\rm AdS}_3}+\frac{2\cos^2\tau}{(1-|w|^2)^2}\Big[(\delta\bar\mu_++\delta\bar\mu_-)dw^2+(\delta\mu_++\delta\mu_-)d\bar w^2\Big]
\cr
&+\frac{2\im\sin\tau\cos\tau}{(1-|w|^2)^2}\Big[(\delta\bar\mu_+-\delta\bar\mu_-)dw^2-(\delta\mu_+-\delta\mu_-)d\bar w^2\Big]
\end{align}
where
\be\label{metric-ads-max}
ds^2_{{\rm AdS}_3}=-d\tau^2+\frac{4\cos^2\tau|dw|^2}{(1-|w|^2)^2}
\ee
is the AdS${}_3$ metric in equidistant coordinates.

Another fact we shall need for our computation is an identity from \cite{Ahlfors:1961} saying that the infinitesimal quasiconformal maps $w+\delta z_\pm$ are area preserving:
\be\nonumber
{\rm Re} \frac{\partial}{\partial w} \frac{\delta z}{(1-|w|^2)^2} = 0.
\ee
Equivalently, by expanding and multiplying by $(1-|w|^2)$, we have
\be\label{ident}
2\frac{\bar w\delta z+w\delta\bar z}{(1-|w|^2)}+\partial_w\delta z+\partial_{\bar w}\delta\bar z=0.
\ee 

Now, to compare the metric (\ref{metric-inf}) arising in the maximal surface description with that in the Fefferman-Graham setting, see below, we could just apply the same
 coordinate transformation that puts the AdS metrics (\ref{metric-ads-max}) into the Fefferman-Graham form to the infinitesimal part of the metric in (\ref{metric-inf}). The
 arising metric could then be expected to be of the Fefferman-Graham type, and one could read off the quantities $a,b$ in terms of the Beltrami coefficients $\delta\mu_\pm$.
 However, this direct approach for performing the computation seems to be too difficult to carry out, and we proceed in a different way.

Thus, let us consider a general vector field
$$\xi=\xi^\tau\partial_\tau+\xi^w\partial_w+\xi^{\bar w}\partial_{\bar w}$$
written in the coordinates relevant for the maximal surface description. We would like to describe the vector fields whose action on the AdS${}_3$ metric (\ref{metric-ads-max})
 gives us the infinitesimal metric (\ref{metric-inf}). We will then be able to compare (asymptotically) such vector field with the Brown-Henneaux vector field generating
 infinitesimal asymptotically AdS metric, see below, and relate the defining parameters in the Brown-Henneaux vector fields with the Beltrami coefficients $\delta\mu_\pm$.
 This will finally lead to a relation between the parameters in the metrics (\ref{metric-inf}) and (\ref{intr-FH}).

Let's start computing the Lie derivative of $ds^2_{{\rm AdS}_3}$ in the direction of $\xi$
\be\nonumber
-2\partial_\tau\xi^\tau d\tau^2+\left(\frac{4\cos^2\tau\partial_\tau\xi^{\bar w}}{(1-|w|^2)^2}-2\partial_w\xi^\tau\right)d\tau dw+\left(\frac{4\cos^2\tau\partial_\tau\xi^w}{(1-|w|^2)^2}-2\partial_{\bar w}\xi^\tau\right)d\tau d\bar w
\cr
+\frac{4\cos^2\tau\partial_w\xi^{\bar w}dw^2}{(1-|w|^2)^2}+\frac{4\cos^2\tau\partial_{\bar w}\xi^wd\bar w^2}{(1-|w|^2)^2}
\cr
+\frac{4\cos^2\tau}{(1-|w|^2)^2}\left(2\frac{\bar w\xi^w+w\xi^{\bar w}}{(1-|w|^2)}+\partial_w\xi^w+\partial_{\bar w}\xi^{\bar w}-2\tan\tau\xi^\tau\right)|dw|^2.
\ee
We now equate this tensor with the infinitesimal part of the metric (\ref{metric-inf}), which leads to the following set of equations:
\be\nonumber
\partial_\tau\xi^\tau=0,\qquad \frac{4\cos^2\tau\partial_\tau\xi^{\bar w}}{(1-|w|^2)^2}-2\partial_w\xi^\tau=0,
\cr
2\frac{\bar w\xi^w+w\xi^{\bar w}}{(1-|w|^2)}+\partial_w\xi^w+\partial_{\bar w}\xi^{\bar w}-2\tan\tau\xi^\tau=0,
\cr
2\partial_{\bar w}\xi^w=(1-\im\tan\tau)\delta\mu_++(1+\im\tan\tau)\delta\mu_-.
\ee

In view of (\ref{inf-beltrami}), the last equation is clearly satisfied by
$$\xi^w=\frac{1}{2}(1-\im\tan\tau)\delta z_++\frac{1}{2}(1+\im\tan\tau)\delta z_-=\frac{1}{2}(\delta z_++\delta z_-)+\frac{1}{2\im}\tan\tau(\delta z_+-\delta z_-)$$
and the third equation becomes
\be\nonumber
2\tan\tau\xi^\tau=\frac{\bar w(\delta z_++\delta z_-)+w(\delta\bar z_++\delta\bar z_-)}{(1-|w|^2)}+\frac{1}{2}\partial_w(\delta z_++\delta z_-)+\frac{1}{2}\partial_{\bar w}(\delta\bar z_++\delta\bar z_-)
\cr
+\tan\tau\left(\frac{1}{\im}\frac{\bar w(\delta z_+-\delta z_-)-w(\delta\bar z_+-\delta\bar z_-)}{(1-|w|^2)}+\frac{1}{2\im}\partial_w(\delta z_+-\delta z_-)-\frac{1}{2\im}\partial_{\bar w}(\delta\bar z_+-\delta\bar z_-)\right).
\ee
Using the identity (\ref{ident}) for each $\delta z_\pm$ we have 
\be\nonumber
\xi^\tau=\frac{1}{2\im}\frac{\bar w(\delta z_+-\delta z_-)-w(\delta\bar z_+-\delta\bar z_-)}{(1-|w|^2)}+\frac{1}{4\im}\partial_w(\delta z_+-\delta z_-)-\frac{1}{4\im}\partial_{\bar w}(\delta\bar z_+-\delta\bar z_-)
\cr
=\frac{1}{\im}\frac{\bar w(\delta z_+-\delta z_-)}{(1-|w|^2)}+\frac{1}{2\im}\partial_w(\delta z_+-\delta z_-).
\ee
Therefore, the generator of the infinitesimal metric (\ref{metric-inf}) has components
\be\nonumber
\xi^\tau_T=\frac{1}{\im}\frac{\bar w(\delta z_+-\delta z_-)}{(1-|w|^2)}+\frac{1}{2\im}\partial_w(\delta z_+-\delta z_-),\cr
\xi^w_T=\frac{1}{2}(\delta z_++\delta z_-)+\frac{1}{2\im}\tan\tau(\delta z_+-\delta z_-).
\ee
This gives us an expression for what can be interpreted as a Brown-Henneaux vector field in the maximal surface description.

Let us now compare this to the holographic description that was sketched in the Introduction. We find it convenient to work with a radial coordinate $\chi=\log(1/\rho)$.
 The metric (\ref{intr-FH}) can then be written as
\be\label{metric-hol}
ds^2=\frac{e^{2\chi}}{4}(-dt^2+d\theta^2)+d\chi^2+\frac{1}{2}(adt^2+2bdtd\theta+ad\theta^2)+\frac{e^{-2\chi}}{4}(a^2-b^2)(-dt^2+d\theta^2).
\ee
The Brown-Henneaux vector fields \cite{Brown:1986nw}, generators of the group of asymptotic symmetries, are parametrized by two functions on the boundary
\begin{align}\label{BH}
\xi^\chi_{BH}=&\;-\frac{1}{2}(\partial_+\xi_++\partial_-\xi_-)+O(e^{-4\chi}),
\cr
\xi^t_{BH}=&\;\frac{1}{2}(\xi_++\xi_-)+e^{-2\chi}(\partial_+^2\xi_++\partial_-^2\xi_-)+O(e^{-4\chi}),
\cr
\xi^\theta_{BH}=&\;\frac{1}{2}(\xi_+-\xi_-)-e^{-2\chi}(\partial_+^2\xi_+-\partial_-^2\xi_-)+O(e^{-4\chi}),
\end{align}
with $\xi_\pm=\xi_\pm(t\pm\theta)$. 

Then, we consider an infinitesimal version of the metric (\ref{metric-hol}) given by
\be\nonumber
ds^2=ds^2_{{\rm AdS}_3}+\frac{1}{2}(\delta adt^2+2\delta bdtd\theta+\delta ad\theta^2),
\ee
where the infinitesimal part is obtained by applying the Lie derivative with respect to the Brown-Henneaux vector field to the AdS${}_3$ metric
$$ds^2_{{\rm AdS}_3}=-\cosh^2\chi dt^2+d\chi^2+\sinh^2\chi d\theta^2,$$
corresponding to $b=0,a=-1$ in (\ref{metric-hol}). The arising relation between the perturbations $\delta a,\delta b$ and the functions $\xi_\pm$ parametrizing the vector field (\ref{BH}) is then given by:
\be\label{var-T}
\delta a=-\frac{1}{2}(\partial_+\xi_++\partial_+^3\xi_+)-\frac{1}{2}(\partial_-\xi_-+\partial_-^3\xi_-),\cr
\delta b=-\frac{1}{2}(\partial_+\xi_++\partial_+^3\xi_+)+\frac{1}{2}(\partial_-\xi_-+\partial_-^3\xi_-).
\ee

We now relate the two descriptions using the fact that they represent different coordinates in the same spacetime. Note that this is indeed possible in a neighbourhood of the
 boundary curve of the maximal surface where both the Fefferman-Graham and the coordinates of (\ref{metric}) are defined. Thus, let us compute the components of the
 Brown-Henneaux vector fields (\ref{BH}) in the coordinates used in (\ref{metric-ads-max}). The coordinate transformation relating the AdS${}_3$ metric in the form
 (\ref{metric-hol}) and its equidistant coordinates description is
\be\nonumber
\tan t=\frac{1-\vert w\vert^2}{1+\vert w\vert^2}\tan\tau,\quad\sinh\chi=\frac{2\vert w\vert}{1-\vert w\vert^2}\cos\tau,\quad \theta=\arg w.
\ee
We can therefore obtain the Brown-Henneaux vector field (\ref{BH}) in equidistant coordinates:
\begin{align}\nonumber
\xi^\tau_{BH}=&\;\frac{1+\vert w\vert^2}{1-\vert w\vert^2}\xi^t_{BH}+\frac{2\vert w\vert\sin\tau}{[(1-\vert w\vert^2)^2+4\vert w\vert^2\cos^2\tau]^{1/2}}\xi^\chi_{BH}
\cr
\xi^w_{BH}=& \;w\tan\tau\xi^t_{BH}-\frac{1}{2}\frac{w}{\vert w\vert}\frac{(1-\vert w\vert^4)\sec\tau}{[(1-\vert w\vert^2)^2+4\vert w\vert^2\cos^2\tau]^{1/2}}\xi^\chi_{BH}+\im w\xi^\theta_{BH}.
\end{align}
We need only the asymptotic behaviour of these components, evaluated on the $\tau=0$ surface. Thus, we get, asymptotically,
\be\nonumber
\xi^\tau_{BH} = \frac{\xi_++\xi_-}{1-|w|^2} + \ldots, \qquad \xi^w_{BH}=\frac{\im w}{2} (\xi_+ -\xi_-) + \ldots,
\ee
where the dots stand for the subleading components.

On the other hand, the maximal surface vector fields (evaluated on the $\tau=0$ surface) are, asymptotically,
\be\nonumber
\xi^\tau_{T}=\frac{1}{\im}\bar w\frac{\delta z_+ -\delta z_-}{1-\vert w\vert^2}+\ldots, \qquad 
\xi^w_{T}=\frac{1}{2}(\delta z_+ +\delta z_-)
\ee
and, equating
\be\nonumber
\xi^\tau_{BH}=\xi^\tau_T, \qquad \xi^w_{BH}=\xi^w_T,
\ee
we immediately read off the sought relations:
\be\label{rel}
\delta z_\pm \Big|_{|w|=1} = \pm \im e^{\im\theta} \xi_\pm.
\ee

\section{The Fefferman-Graham Stress Tensor and the Bers Embedding}
\label{sec:bers}

We now develop an interpretation of the obtained relation (\ref{rel}) between the maximal surface and the holographic descriptions. We remain at the infinitesimal level,
 leaving the question of explicitly relating the general metric from evolving data (\ref{I}) and (\ref{II}) and the Fefferman-Graham metric (\ref{metric-hol}) to future studies.

So far, we have seen how the two functions $\xi_\pm$ parametrizing the Brown-Henneaux vector fields are given in terms of the boundary values of the (infinitesimal) quasiconformal
 maps $\delta z_\pm$. However, a more interesting question is that of a relation between the holographic stress-energy tensor components --- functions $a,b$ in (\ref{metric-hol})
 --- and these quasiconformal maps. In this Section we shall see that this relation is that between the holomorphic quadratic differentials arising via the so-called Bers embedding
 of $\T(1)$ and the quasiconformal maps parametrizing $\T(1)$. In other words, we shall see that the stress-energy tensor components of the holographic description are nothing but
 the components of the quadratic differentials that arise via the Bers embedding of $\T(1)\times\T(1)$. 

Let us start by reminding the reader some facts about the Bers embedding. We are necessarily brief here, and for more details the reader can consult a very accessible exposition
 in \cite{Nag:1993}. The Bers embedding arises via the so-called B-model of the universal Teichm\"uller space. So far we have been solving the Beltrami equation
$$\partial_{\bar{w}} z = \mu \partial_w z$$
 for a quasiconformal map $z_\mu$ starting from a (bounded) Beltrami coefficient $\mu$ on the unit disc $\Delta$, and then extending $\mu$ symmetrically to its complement
 $\Delta^\ast$. This gives rise to a quasiconformal map $z_\mu$, depending real-analytically on $\mu$, that leaves the unit disc invariant (and thus also the unit circle
 and the complement of the disc), and reduces to a quasisymmetric map on the unit circle. This gave rise to the A-model of the universal Teichm\"uller space.

The B-model is obtained by taking the same bounded Beltrami coefficients on $\Delta$, and then extending them to be zero on $\Delta^\ast$. The corresponding quasiconformal
 map $z^\mu$ is now conformal on $\Delta^\ast$ , it is in fact biholomorphic onto its image, and depends on $\mu$ complex-analytically. The universal Teichm\"uller space in
 the B-model is defined as the space of equivalence classes of Beltrami coefficients (or quasiconformal maps), with two coefficients considered equivalent if the corresponding
 solutions agree outside the unit disc. Thus $\T(1)$ in the B-model is parametrized by univalent holomorphic functions on $\Delta^\ast$ and one may use the holomorphicity of
 these maps to obtain an embedding of $\T(1)$ in the space of holomorphic quadratic differentials on the complement of $\Delta$ via Schwarzian derivative.

At the infinitesimal level, there exists an explicit relation between the two models exhibited in \cite{Nag:1993}. Thus, the tangent vectors to $\T(1)$ in the A-model can be
 described as functions $u:\bbS^1\rightarrow\R$ on the circle that are related to the so-called Zygmund class functions on the real line, see \cite{Nag:1993} and references
 therein for more details. What is important for us here is that the functions $u$ are defined from the boundary values of the corresponding infinitesimal quasiconformal maps
 $\delta z$ via:
\be\label{u}
u(e^{\im\theta})=\frac{\delta z(e^{\im\theta})}{\im e^{\im\theta}}.
\ee
The functions arising as tangent vectors to $\T(1)$ are those for which 
\be\nonumber
F(x) = \frac{1}{2} (x^2+1) u\left( \frac{x-\im}{x+\im}\right)
\ee
are of the Zygmund class, see \cite{Nag:1993}. The functions $u$ can be expanded into a Fourier series:
\be\nonumber
u(e^{\im\theta}) = \sum_{k=-\infty}^{\infty} u_k e^{\im k\theta}
\ee
where $u_{-k}=\bar u_k$, as required by the fact $u(e^{\im\theta})$ is a real function.

The tangent space to the B-model universal Teichm\"uller space is described as follows. In this model the solution $z^\mu$ of the Beltrami equation is of the form
 $z^\mu(w)=w+\delta z(w)$, with the function $\delta z$ now being holomorphic on $\Delta^\ast$. It thus admits an expansion
\be\label{dz-B}
w+\delta z(w)= w\left( 1+\frac{c_2}{w^2}+\frac{c_3}{w^3}+\ldots \right) \qquad {\rm in}\quad |w|>1,
\ee
where a M\"obius transformation is used to remove the $1/w$ term in the brackets and to set the first term to unity. The corresponding holomorphic quadratic differential,
 obtained as the (infinitesimal) Schwarzian derivative $\partial_w^3\delta z$ of (\ref{dz-B}), can also be expanded as
\be\label{h}
h(w) = \frac{1}{w^4}\left(h_0 + \frac{h_1}{w} + \frac{h_2}{w^2} +\ldots \right) \qquad {\rm in}\quad |w|>1,
\ee
where the coefficients $h_k$ are related to those in (\ref{dz-B}) via
\be\label{h-c}
h_{k-2} = c_k (k-k^3), \qquad k\geq 2.
\ee
Finally, the relation between the A- and B-model Fourier coefficients $u_k$ and $c_k$ is given by, see \cite{Nag:1993}
\be\label{c-a}
c_k =\im \bar u_k.
\ee

For the later purposes, we now note that in all the discussions of the B-model above we could have replaced $\Delta$ and $\Delta^\ast$. In fact, this is the choice made in
 some of the references, see e.g. \cite{Takhtajan:2006}. In this case one works with bounded Beltrami differentials on $\Delta^\ast$, solves the Beltrami equation continuing
 $\mu=0$ inside, and gets a univalent holomorphic function on $\Delta$, whose Schwarzian derivative produces a holomorphic quadratic differential on the unit disc. We could
 have as well worked with these models for the universal Teichm\"uller space. In fact, as we shall see below, it will be natural to work with holomorphic functions on $\Delta^\ast$
 for one copy of $\T(1)$ and with holomorphic functions on $\Delta$ for the other. The analogues of (\ref{dz-B}) and (\ref{h}) in this realization of $\T(1)$ are then given by
\be\nonumber
w+\delta \hat{z}(w) = w(1+\hat{c}_2 w^2 + \hat{c}_3 w^3+\ldots ), \qquad |w|<1,
\ee
and
\be\label{h-1}
\hat{h}(w) = \hat{h}_0 + \hat{h}_1 w+ \hat{h}_2 w^2 +\ldots \qquad |w|<1,
\ee
where we have denoted the quantities arising in this realization of the B-model by letters with an extra hat. We also note that there is an extra minus as compared to (\ref{h-c})
 in the relation between the coefficients in this realization
$$\hat{h}_{k-2}=-\hat{c}_k (k-k^3).$$
Also, note that we can always map a holomorphic function inside the disc to an anti-holomorphic function outside by $w\to 1/\bar{w}$. By applying this to the quadratic
 differential (\ref{h-1}) we get a new anti-holomorphic quadratic differential $h(\bar w)$ outside of the disc by complex conjugating
\be\label{h-2}
\overline{h(\bar{w})}=\hat{h}\left(\frac{1}{w}\right) \frac{1}{w^4} = \frac{1}{w^4}\left(\hat{h}_0 + \frac{\hat{h}_1}{w} + \frac{\hat{h}_2}{w^2} +\ldots \right) \qquad {\rm in}\quad |w|>1,
\ee
which gives the same expression as in (\ref{h}), but with the change $w\to\bar{w}$. We shall use this realization in terms of anti-holomorphic functions on $\Delta^\ast$
 for the second copy of $\T(1)$, and the ``usual'' realization in terms of holomorphic functions on $\Delta^\ast$ for the first copy of the universal Teichm\"uller.

Another relation we shall need is that between the $c$-coefficients in the realization of the B-model in terms of Beltrami coefficients on $\Delta^\ast$ (we have denoted these
 coefficients by $\hat{c}_k$ above), and the Fourier coefficients $u_k$ of the Zygmund functions $u$ of the A-model. Formula (\ref{c-a}) gives such relation for one realization
 of the B-model, and we need to derive a similar relation for the other. The derivation is a straightforward adaptation of the Proof I in \cite{Nag:1993}. One finds
\be\label{c-hat}
\hat{c}_k = -\frac{1}{\pi} \int_{\Delta^*} \frac{\mu(w)}{w^{k+2}} dx\,dy.
\ee
In relating it to the A-model Fourier coefficients we use the fact that the A-model Beltrami coefficient on $\Delta^\ast$ is obtained by reflection. Thus, the integral in
 (\ref{c-hat}) can be taken over the unit disc with the reflected Beltrami given by
\be\nonumber
\mu\left(\frac{1}{\bar{w}}\right) = \overline{\mu(w)} \frac{w^2}{\bar{w}^2}.
\ee
Substituting this to the integral, and taking into account the change of integration measure $dx\, dy\to - dx\, dy / w^2 \bar{w}^2$ we get
\be\nonumber
\hat{c}_k = \frac{1}{\pi} \int_{\Delta} \overline{\mu(w)} \bar{w}^{k-2} dx\,dy.
\ee
The hatted $c$-coefficients are thus related to the unhatted ones by complex conjugation
\be\nonumber
\hat{c}_k = \bar c_k.
\ee

We can now continue to think about both points in $\T(1)\times \T(1)$ as being parametrized by Beltrami coefficients inside $\Delta$. This gives two A-model quasiconformal
 maps in $\Delta$ whose boundary values produce two Zygmund functions $u$. Expanding these into Fourier modes we get two sets of coefficients $u_k$, which we shall later denote
 by $u_k^\pm$. Now our convention is that the  B-model for the first copy of $\T(1)$ is obtained by setting to zero the Beltrami on the complement of $\Delta$, thus producing
 holomorphic functions on $\Delta^*$, and the holomorphic quadratic differentials as in (\ref{h}). The B-model for the second copy of $\T(1)$ will be realized by first reflecting
 the Beltrami coefficient, then setting it to zero inside $\Delta$, thus producing a holomorphic quadratic differential on $\Delta$, which in turn can be interpreted as an
 anti-holomorphic quadratic differential on $\Delta^*$, as in (\ref{h-2})

Collecting all the relations above we can write the following relations between the $u$- and $h$-coefficients of the two copies of $\T(1)$:
\be\label{h-u-pm}
h^\pm_{k-2} = \pm\im\bar u^\pm_k(k-k^3), 
\ee
where we have now differentiated between the two copies of $\T(1)$ by assigning plus and minus labels. We also used the following notation in (\ref{h-2}) relating the hatted $h$ coefficients and the coefficients of the anti-holographic quadratic differential:
$$\hat{h}^-_{k-2}=\bar h^-_{k-2}.$$

Now, to prepare for the relation between the Bers embedding quadratic differentials and the holomorphic stress-energy tensor that we will derive, let us rewrite the stress tensor 
\be\nonumber
T= adt^2 + 2b dt d\theta + a d\theta^2
\ee
of the Fefferman-Graham metric in a more suggestive way. To this end, we will analytically continue the $t$ coordinate to the imaginary values. Thus, let us continue all the
 functions appearing in $T$ via
\be\label{an-cont}
t = \frac{1}{2\im} \log|w|^2, \qquad \theta = \frac{1}{2\im} \log\frac{w}{\bar{w}},
\ee
so that the new (imaginary) time coordinate runs between $-\im\infty$ and $\im \infty$ while $w$ runs over the complex plane with the unit circle $|w|=1$ corresponding to $t=0$.
 With this choice we have
\be\nonumber
t+\theta = \frac{1}{\im}\log w, \qquad t-\theta = \frac{1}{\im}\log \bar{w},
\ee
so that functions of $e^{\im (t\pm\theta)}$ become holomorphic (anti-holomorphic) functions on the complex plane. In particular, the functions $a_\pm$ whose sum and difference give $a,b$
 would seem to become a holomorphic and anti-holomorphic function on the complex plane. However, there is no (bounded) holomorphic function on the whole complex plane apart from
 a constant. Thus, we need to be very careful when designing the analytic continuation.

Let us expand $a_\pm$ into Fourier modes. When restricted to $t=0$ these are periodic functions of $\theta$, and so the Fourier expansion is possible. We have
\be\nonumber
a_\pm(t\pm \theta) = \sum_{k=-\infty}^\infty a_k^\pm e^{\im k (t\pm\theta)}.
\ee
As before, we have $\bar a^\pm_k=a^\pm_{-k}$ so that these are real functions. It is clear that we cannot continue $a_\pm$ as holomorphic or anti-holomorphic functions into the
 whole complex plane. What is possible is to take what can be called the chiral part of $a_\pm$, containing only, say, the negative frequency modes, and continue this part only.
 Thus, let us introduce
\be\nonumber
\tilde{a}_\pm(t\pm \theta) = \sum_{k=-\infty}^{-2} a_k^\pm e^{\im k (t\pm\theta)}
\ee
for the chiral parts. Here we have used the fact that $|k|\geq 2$ in these expansions, which will become manifest below. We now continue the chiral parts via (\ref{an-cont}) to
 the complement of the unit disc to get:
\be\label{a-pm-1}
\tilde{a}_+(w) = \frac{a_{-2}^+}{w^2} + \frac{a_{-3}^+}{w^3} + \ldots, \qquad
\tilde{a}_-(\bar{w}) = \frac{a_{-2}^-}{\bar{w}^2} + \frac{a_{-3}^-}{\bar{w}^3} + \ldots,
\ee
which are, respectively, holomorphic and anti-holomorphic functions on $\Delta^\ast$. We can now analytically continue the chiral part $\tilde T$ of the stress tensor
 (\ref{h-1}), which is the tensor $T$ with functions $a_\pm$ replaced by their chiral parts. A simple computation gives:
\be\nonumber
\tilde{T} =  - \frac{\tilde{a}_+(w)}{w^2} dw^2-\frac{\tilde{a}_-(\bar{w})}{\bar{w}^2} d\bar{w}^2.
\ee

We may now relate the chiral parts $\tilde{a}_\pm$ of the stress-energy tensor to the quadratic differentials arising via the Bers embedding of $\T(1)\times \T(1)$. To this
 end, let us first obtain the relation between the Fourier coefficients $a^\pm_k$ and those of the Fourier expansions of the functions $\xi_\pm$ appearing in the Brown-Henneaux
 vector fields. From (\ref{var-T}) we have
\be\nonumber
-2 a_\pm=\partial_\pm \xi_\pm +\partial_\pm^3 \xi_\pm.
\ee
Thus, if we expand
\be\nonumber
\xi_\pm(t\pm\theta) = \sum_{k=-\infty}^{\infty} \xi^\pm_k e^{\im k(t\pm\theta)},
\ee
with $\xi^\pm_{-k}=\bar\xi^\pm_k$, we get the following relation between the Fourier coefficients:
\be\nonumber
-2 a^\pm_k = \im\xi^\pm_k(k-k^3).
\ee

We now come back to the problem of relating the maximal surface and holographic descriptions. We have seen that the relation between the Brown-Henneaux functions $\xi_\pm$ and
 the boundary values of the quasiconformal maps $\delta z_\pm$ is given by (\ref{rel}). In view of (\ref{u}), this can then be rewritten as
\be\nonumber
\xi_\pm(\pm \theta) = \pm u^\pm (e^{\im\theta}),
\ee
where on the left-hand-side the functions $\xi_\pm$ are restricted to the circle $t=0$. This now implies the following relation between the Fourier coefficients of the
 $\xi$-functions and the A-model functions $u_\pm$ 
\be\nonumber
\xi^\pm_k = \pm u^\pm_k,
\ee
with $u^\pm_{-k} = \bar u^\pm_k$. We can therefore write $-2 a_k^\pm=\pm \im u_k^\pm (k-k^3)$ or
\be\nonumber
2a_{-k}^\pm = \pm \im \bar u_k^\pm(k-k^3).
\ee
Comparing this with (\ref{h-u-pm}) we see that
\be\nonumber
h^\pm_{k-2} = 2a_{-k}^\pm, \qquad k\geq 2.
\ee
This gives us the desired relation between the coefficients in the expansions (\ref{a-pm-1}) of the chiral parts of the stress-energy tensor and those of the Bers embedding quadratic
 differentials (\ref{h}), (\ref{h-2}), and implies
\be\label{a-h}
\frac{2\tilde{a}_+(w)}{w^2} = h^+(w), \qquad \frac{2\tilde{a}_-(\bar{w})}{\bar{w}^2} =h^-(\bar{w}).
\ee
Finally, the analytic continuation of the chiral part $\tilde{T}$ of the stress-energy tensor is equal to (minus half) the sum of two quadratic differentials arising via the Bers
 embedding:
\be\label{T-h}
\tilde{T}= -\frac{1}{2} h^+(w) dw^2 -\frac{1}{2}h^-(\bar{w}) d\bar{w}^2,
\ee
which is our final result for the (infinitesimal) relation between the maximal surface and the holographic descriptions.

\section{Charges}
\label{sec:charges}

Having the relation (\ref{T-h}) we now wish to derive an expression for the asymptotic charges for a general spacetime in terms of the maximal surface parametrization. Similar
 to the previous Section, we shall do so at the infinitesimal level. However, the answer that we obtain admits an obvious generalization to the finite case. We shall see that
 the charges are given simply by the (real parts of) the periods of the Bers embedding quadratic differentials. 

We start by computing the charges of an asymptotically AdS spacetime in the holographic Fefferman-Graham description, see \cite{Balasubramanian:1999}. Here one starts with
 the Einstein-Hilbert action complemented with the York-Gibbons-Hawking boundary term and a volume renormalization counterterm
\be\nonumber
S=\frac{1}{2}\int_M d^3x\sqrt{-g}(R-2\Lambda)+\int_{\partial M} d^2x\sqrt{-\gamma}(\Theta-1)
\ee
The quasi-local stress-energy tensor is then obtained from the variation of the action with respect to the boundary metric
\be\nonumber
T_{\mu\nu}=\frac{2}{\sqrt{-\gamma}}\frac{\delta S}{\delta \gamma^{\mu\nu}}.
\ee
One gets
\be\nonumber
T_{\mu\nu}=-\Big(\Theta_{\mu\nu}-\Theta^\rho{}_\rho \gamma_{\mu\nu}+\gamma_{\mu\nu}\Big)
\ee
where $\gamma$ is the boundary metric and $\Theta$ the boundary extrinsic curvature. 

Let us now consider a spacetime asymptotically described by a Fefferman-Graham type metric (\ref{metric-hol}). Let $M_{\chi_0}$ denote the portion of the spacetime
 manifold where $\chi<\chi_0$. The metric induced on $\chi=\chi_0$ surface is
\be\nonumber
\gamma=\frac{1}{4}e^{2\chi_0}(-dt^2+d\theta^2)+\frac{1}{2}(adt^2+2bdtd\theta+ad\theta^2)+\frac{1}{4}e^{-2\chi_0}(a^2-b^2)(-dt^2+d\theta^2).
\ee
The extrinsic curvature of the $\chi=\chi_0$ surface is given by
\be\nonumber
\Theta_{\mu\nu}=-\mathcal{L}_n\gamma_{\mu\nu}\vert_{\chi_0}=-\nabla_\mu n_\nu\vert_{\chi_0}=-\frac{1}{2}\partial_\chi g_{\mu\nu}\vert_{\chi_0}
\ee
where $n=-\partial_\chi$ is the unit normal vector field to the boundary $\partial M_{\chi_0}$.
Therefore
\be\nonumber
\Theta=\frac{1}{4}\left[e^{2\chi_0}-e^{-2\chi_0}(a^2-b^2)\right](-dt^2+d\theta^2)
\ee
and, taking the limit $\chi_0\to\infty$, we get
\be\nonumber
T=adt^2+2bdtd\theta+ad\theta^2.
\ee

Then for each asymptotic Killing vector field $\xi$ we have a conserved charge given by
\be\nonumber
Q_{\xi}=\lim_{\chi_0\rightarrow\infty}\frac{1}{2\pi}\int_{\partial\Sigma_{\chi_0}}d\theta\sqrt{\vert\sigma\vert}u^\mu\xi^\nu T_{\mu\nu}.
\ee
Here we need to consider a spacelike slice $\Sigma=\{t=0\}$ and compute its unit normal timelike vector field
\be\nonumber
u=\frac{2}{e^{2\chi}-e^{-2\chi}(a^2-b^2)}\left[\sqrt{e^{2\chi}+2a+e^{-2\chi}(a^2-b^2)}\partial_t-\frac{2b}{\sqrt{e^{2\chi}+2a+e^{-2\chi}(a^2-b^2)}}\partial_\theta\right]
\ee
We also consider $\Sigma_{\chi_0}=M_{\chi_0}\cap\Sigma$. The induced metric on its boundary $\partial\Sigma_{\chi_0}$ is
\be\nonumber
\sigma=\frac{1}{4}\left[e^{2\chi_0}+2a+e^{-2\chi_0}(a^2-b^2)\right]d\theta^2.
\ee
Mass is the conserved charge associated with time translation ($\xi=\partial_t$)
\be\nonumber
M=\lim_{\chi_0\rightarrow\infty}\frac{1}{2\pi}\int_{\partial\Sigma_{\chi_0}}d\theta\sqrt{\vert\sigma\vert}(u^0 T_{00}+u^2 T_{20})=\frac{1}{2\pi}\int_{\partial\Sigma}d\theta\,a(\theta)
\ee
and angular momentum the one associated with rotation ($\xi=-\partial_\theta$)
\be\nonumber
J=\lim_{\chi_0\rightarrow\infty}\frac{1}{2\pi}\int_{\partial\Sigma_{\chi_0}}d\theta\sqrt{\vert\sigma\vert}(u^0 T_{02}+u^2 T_{22})=\frac{1}{2\pi}\int_{\partial\Sigma}d\theta\,b(\theta).
\ee
Taking into account the relations
$$a(t,\theta)=a_+(t+\theta)+a_-(t-\theta), b(t,\theta)=a_+(t+\theta)-a_-(t-\theta)$$
to the chiral functions $a_\pm$ we can rewrite the above formulas for the charges compactly as
\be\nonumber
\frac{1}{2}(M\pm J) = \frac{1}{2\pi} \oint_{\partial\Sigma} a_\pm.
\ee
Note that, even prior to any relation to the maximal surface description, these can be expressed as the real parts of the periods of the holomorphic quadratic differentials
 arising via the analytic continuation of the chiral parts of $a_\pm$, see the previous Section. 

We now relate this to the maximal surface description. From (\ref{a-h}) we know that the chiral parts of the functions $a_\pm$ are basically the (analytic continuations
 of the) quadratic differentials $h_\pm$ arising from the Bers embedding. The full functions $a_\pm$ on the circle can be obtained by taking their chiral parts and adding
 the complex conjugate. Thus, we have $2{\rm Re} (\tilde{a}^\pm)|_{|w|=1} = a^\pm(\theta)$ and therefore
\be\label{charges}
\frac{1}{2}(M+J)=\frac{1}{2\pi} {\rm Re} \oint_{|w|=1} w^2 h^+ , \qquad
\frac{1}{2}(M-J)=\frac{1}{2\pi}{\rm Re} \oint_{|w|=1} \bar w^2 h^-.
\ee
which are just the (real parts of the) periods of the Bers embedding quadratic differentials $h^\pm$. The formulas in terms of the Bers embedding quadratic differentials are
 of course only valid at the infinitesimal level, where we have a relation between the functions $a_\pm$ of the holographic description and the data on the maximal surface.
 However, as we noted above, the same formulas are valid even in the finite case if one understands that $h_\pm$ are the (multiples of the) analytic continuations of the chiral
 parts of $a_\pm$, see (\ref{a-h}). It is then natural to conjecture that the analytic continuations of the chiral parts of $a_\pm$ continue to be related to the Bers embedding
 quadratic differentials in the same way as they do in the infinitesimal case, and that (\ref{charges}) gives a general formula for the charges in terms of the maximal surface
 data. We leave an attempt at demonstration this finite case relation to future work. We also note that in this infinitesimal case the charges (\ref{charges}) are actually zero,
 for there is no $1/w^2, 1/\bar{w}^2$ terms in the expansion of the infinitesimal quadratic differentials $h^\pm$, see (\ref{h}), (\ref{h-2}). So, the first order variation of
 the charges, computed at the origin corresponding to the AdS${}_3$ is zero. It is clear however that considering non-trivial spatial topologies, obtained as the quotients of
 ${\rm AdS}_3$ by some discrete groups of isometries, will render non-trivial periods for the (anti-)holomorphic quadratic differentials and therefore non-trivial charges in
 each asymptotic region.

\section{The Phase Space Symplectic Structure}
\label{sec:sympl}
We turn to the description of the gravitational symplectic structure on the universal phase space. First, we must warn the reader that, due to noncompactness of $\Delta$, this
 is only a formal symplectic structure. In fact, already in the universal Teichm\"uller space context the universal Weil-Petersson symplectic structure (in fact the Weil-Petersson
 hermitian metric) diverges for certain tangent directions. A solution for this problem was introduced in \cite{Takhtajan:2006} where a new topology is introduced in $\T(1)$ which
 makes it into a Hermitian manifold with a well defined Weil-Petersson hermitian metric in each tangent space. We shall not reproduce their arguments here and refer the reader to
 \cite{Takhtajan:2006} for more information. We note however that the results of \cite{Takhtajan:2006} directly extend to the universal phase space and, therefore, the formal results
 obtained below can be readily made more rigorous.

We first compute the symplectic structure in the cotangent bundle description, and then translate it into the generalized Mess description $\T(1)\times \T(1)$ using the Mess map.
 We shall see that the pull-back of the canonical cotangent bundle symplectic structure on $T^\ast \T(1)$ to $\T(1)\times \T(1)$ coincides with the difference of Weil-Petersson
 symplectic structures coming from each copy of $\T(1)$. Thus, the generalized Mess map is symplectic. We do computations by comparing the symplectic structures at the origin of
 both spaces. The result at an arbitrary point should then follow using the group structure of the universal Teichm\"uller space $\T(1)$, but we shall not attempt to demonstrate
 this in the present paper. Instead, next Section shows the Chern-Simons $SL(2,\R)$ connections are respectively parametrized by $\mu_\pm$, which then implies the gravitational
 symplectic structure should indeed coincide with the difference of Weil-Petersson symplectic structures in each sector of the theory.

From the Hamiltonian formulation of general relativity, one knows that the pre-symplectic 1-form is given by
$$\Theta=\frac{1}{2}\int_\Delta(\II,\delta I)_Ida_I=\frac{1}{2}\int_\Delta\tr(I^{-1}\II I^{-1}\delta I)da_I.$$
When working in the maximal surface with its first and second fundamental forms given by $$I=e^{2\varphi}|dz|^2,\quad\II=\frac{1}{2}(qdz^2+\bar qd\bar z^2),$$
we have for the first variation of $I$:
$$\delta I=e^{2\varphi}\big(\delta\bar\mu dw^2+\delta\mu d\bar w^2+(2\delta\varphi+\partial_w\delta z+\partial_{\bar w}\delta\bar z)|dw|^2\big).$$
Here $\mu$ is the Beltrami differential describing variations of the conformal structure of the maximal surface. The pre-symplectic 1-form is therefore
$$\Theta=\int_\Delta d^2w(q\delta\mu+\bar q\delta\bar\mu).$$
We see that the holomorphic quadratic differential determining the second fundamental form is canonically conjugated to the variable $\mu$ parametrizing the conformal
 structure of the maximal surface. We note that it is the same computation that is valid in the context of compact spatial sections AdS${}_3$ manifolds and in our context
 of asymptotically AdS${}_3$ spacetimes. Taking the variation of the pre-symplectic 1-form we get
$$\Omega=\int_\Delta d^2w(\delta q\wedge\delta\mu+\delta \bar q\wedge\delta\bar\mu),$$
which shows that the symplectic structure induced by the Einstein-Hilbert functional is just the canonical cotangent bundle symplectic structure on $T^\ast \T(1)$.

Now, using the Mess map we can write the variations of $\mu$ and $q$ at the origin (corresponding to AdS${}_3$) in terms of those of $\mu_\pm$
$$\delta\mu=\frac{1}{2}(\delta\mu_++\delta\mu_-),\qquad\delta q=\frac{4\im}{(1-|w|^2)^2}\delta\bar\nu=\frac{2\im}{(1-|w|^2)^2}(\delta\bar\mu_+-\delta\bar\mu_-).$$
The gravitational symplectic form, evaluated at the origin of the phase space, therefore becomes
\be\label{WP-S}
\Omega=\frac{1}{2\im}\int_\Delta\frac{4d^2w}{(1-|w|^2)^2}(\delta\mu_+\wedge\delta\bar\mu_+-\delta\mu_-\wedge\delta\bar\mu_-),
\ee
which is just a copy of the Weil-Petersson symplectic form in each $\T(1)$. This shows the generalized Mess map $T^\ast \T(1)\to \T(1)\times \T(1)$ (at the origin of both
 spaces) is symplectic. It should be possible to extend this to an arbitrary point by using the group structure of the universal Teichm\"uller space, see the Appendix, but
 we shall not attempt this here.

\section{Chern-Simons connections}
\label{sec:CS}

Finally, in this Section, we present a relation between the $\T(1)\times\T(1)$ parametrization of the phase space and Chern-Simons formulation of 2+1 general relativity \cite{Witten:1988hc}. We remind the reader that in the first order formalism the variables one works with are a frame field $e$ and a spin connection $\omega$. For negative cosmological constant, these may be combined into a ${\rm SL}(2,\R)\times {\rm SL}(2,\R)$ connection $A=(A^+,A^-)$ over spacetime $M=\R\times\Delta$
\be\nonumber
A^{\pm}_{\mu}=\left(\omega^a_{\mu}\pm e^a_{\mu}\right)T_a\,,
\ee
where $T_a$ are the generators of ${\rm SL}(2,\R)$. Here we choose to work with ${\rm SU}(1,1)$ generators
\be\nonumber
T_0=\frac{\im}{2}\left[\begin{matrix} 1 & 0 \cr 0 & -1 \end{matrix}\right],\quad T_1=\frac{1}{2}\left[\begin{matrix} 0 & -1 \cr -1 & 0\end{matrix}\right],\quad T_2=\frac{1}{2}\left[\begin{matrix} 0 & -\im \cr \im & 0 \end{matrix}\right]
\ee
so that we have
\be\nonumber
\tr(T_aT_b)=\frac{1}{2}\eta_{ab},\qquad[T_a,T_b]=\epsilon_{ab}{}^cT_c.
\ee

Written in terms of $A^+$ and $A^-$, the Einstein-Hilbert action becomes the difference of two Chern-Simons action
\be\nonumber
S_{EH}[A^+,A^-]=S_{CS}[A^{+}]-S_{CS}[A^{-}].
\ee
In this sense we say that 2+1 GR is equivalent to Chern-Simons theory. Note, however, that the phase space of Chern-Simons theory, which is the space of all solutions of
 the equations of motion, is much bigger than that of GR. Basically, some connections define non-invertible frames $e$ so that singular metrics are also included. The gauge
 group of CS theory also includes some transformations (large gauge transformations) that cannot be considered as gauge from the point of GR. In spite of this, the CS point
 of view on AdS${}_3$ gravity is very convenient, because it gives the simplest way to understand how the Mess-type description by two copies of the Teichm\"uller space can
 be possible.

The CS formulation thus shows there exists a pair of flat ${\rm SL}(2,\R)$ connections associated with any AdS metric. To relate our phase space construction to CS theory we
 compute the flat ${\rm SL}(2,\R)$ connections associated with the AdS metric parametrized by $(f_+,f_-)$. Again, it is convenient to start working at the maximal surface.
 The 3-metric can then be written
\be\nonumber
ds^2=-d\tau^2+\cos^2\tau e^{2\varphi}\vert dz\vert^2+\sin\tau\cos\tau(qdz^2+\bar qd\bar z^2)+\sin^2\tau e^{-2\varphi}|q|^2\vert dz\vert^2,
\ee
and it is a simple computation to find the associated flat ${\rm SL}(2,\R)$ connections
\be\nonumber
A^\pm_z=\frac{1}{2}\left[\begin{matrix} \partial_z\varphi & \mp e^{\varphi} \cr \im e^{-\varphi}q & -\partial_z\varphi\end{matrix}\right],\quad A^\pm_{\bar z}=\frac{1}{2}\left[\begin{matrix} -\partial_{\bar z}\varphi & -\im e^{-\varphi}\bar q \cr \mp e^{\varphi} & \partial_{\bar z}\varphi \end{matrix}\right].
\ee
Here we have eliminated the $\tau$ dependence using a gauge transformation. Recalling that the Liouville field and the holomorphic quadratic differential can be written,
 in terms of $F_\pm$, as
$$e^{2\varphi}=\frac{4|\partial F_+|^2}{(1-|z|^2)^2},\qquad qdz^2=\im{\rm Hopf}(F_+),$$
we need just another gauge transformation $a_\pm\rightarrow g^{-1}a_\pm g+g^{-1}dg$, with
\be\nonumber
g=\left[\begin{matrix} \left(\partial_z F_\pm/|\partial_z F_\pm|\right)^{-1/2} & 0 \cr 0 & \left(\partial_z F_\pm/|\partial_z F_\pm|\right)^{1/2} \end{matrix}\right],
\ee
to see the connections decouple. A pull-back to the base disc then gives us
\be\nonumber
A^\pm_w=\frac{1}{(1-|z_\pm|^2)}\left[\begin{matrix} \frac{1}{2}(\bar z_\pm\partial_{w}z_\pm-z_\pm\partial_{w}\bar z_\pm) & \mp\partial_{w}z_\pm \cr \mp\partial_{w}\bar z_\pm & -\frac{1}{2}(\bar z_\pm\partial_{w}z_\pm-z_\pm\partial_{w}\bar z_\pm)\end{matrix}\right],
\ee
\be\nonumber
A^\pm_{\bar w}=\frac{1}{(1-|z_\pm|^2)}\left[\begin{matrix} -\frac{1}{2}(\bar z_\pm\partial_{\bar w}z_\pm-z_\pm\partial_{\bar w}\bar z_\pm) & \mp\partial_{\bar w}\bar z_\pm \cr \mp\partial_{\bar w}z_\pm & \frac{1}{2}(\bar z_\pm\partial_{\bar w}z_\pm-z_\pm\partial_{\bar w}\bar z_\pm)\end{matrix}\right],
\ee
and we see that each copy of $\T(1)$ parametrizes one of the CS sectors, as expected.

We may then compute the Chern-Simons pre-symplectic structure in this parametrization. Let's work with a single $SL(2,\R)$
Chern-Simons theory on $\R\times\Delta$ for the moment. We again compute the symplectic structure at the base point in $\T(1)$. Then, a flat $SL(2,\R)$ connection is
 simply given by
\be\nonumber
A=\frac{1}{(1-|w|^2)}\left[\begin{matrix} \frac{1}{2}(\bar wdw-wd\bar w) & -dw \cr -d\bar w & -\frac{1}{2}(\bar wdw-wd\bar w)\end{matrix}\right].
\ee
Its variation in the direction of a tangent vector $\delta\mu\in T_{[0]}\T(1)$ is then easily obtained
\be\nonumber
\delta A_w=\frac{1}{2}\left[\begin{matrix} -\frac{1}{2}\partial_w\Big(\partial_w\delta w-\partial_{\bar w}\delta\bar w\Big) & -\frac{\partial_w\delta w-\partial_{\bar w}\delta\bar w}{(1-|w|^2)} \cr -\frac{2\delta\bar\mu}{(1-|w|^2)} & \frac{1}{2}\partial_w\Big(\partial_w\delta w-\partial_{\bar w}\delta\bar w\Big)\end{matrix}\right],
\ee
\be\nonumber
\delta A_{\bar w}=\frac{1}{2}\left[\begin{matrix} -\frac{1}{2}\partial_{\bar w}\Big(\partial_w\delta w-\partial_{\bar w}\delta\bar w\Big) & -\frac{2\delta\mu}{(1-|w|^2)} \cr \frac{\partial_w\delta w-\partial_{\bar w}\delta\bar w}{(1-|w|^2)} & \frac{1}{2}\partial_{\bar w}\Big(\partial_w\delta w-\partial_{\bar w}\delta\bar w\Big)\end{matrix}\right].
\ee
Here we made use of identity (\ref{ident}) as well as
$$\partial_w\delta\mu=-\frac{2\bar w\delta\mu}{(1-|w|^2)},$$
which follow directly from the representation of Harmonic Beltrami coefficients in terms of holomorphic quadratic differentials.

The pre-symplectic structure is then obtained by restriction from the natural symplectic structure on the space of all connections,
\be\nonumber
\Omega_{CS}=\int_S\tr(\delta A\wedge\delta A)=\im\int_Sd^2w\tr(\delta A_w\wedge\delta A_{\bar w}).
\ee
Up to a boundary term, this is simply the Weil-Petersson symplectic structure
\be\nonumber
\Omega_{CS}=\im\int_Sd^2w\frac{\delta\bar\mu\wedge\delta\mu}{(1-|w|^2)^2}+\frac{1}{2}\int_{\partial S}(\partial_w\delta w-\partial_{\bar w}\delta\bar w)\wedge d(\partial_w\delta w-\partial_{\bar w}\delta\bar w)
\ee
in agreement with the results of the previous Section.

\section{Discussion}

In this paper we described an explicit parametrization of a large class of AdS${}_3$ manifolds by two copies of the universal Teichm\"uller space $\T(1)$. Our construction
 proceeds by first determining the first and second fundamental forms on the maximal surface that corresponds to a given point in $\T(1)\times \T(1)$, and then evolving this
 initial data using (\ref{metric}) to get the spacetime metric. We note that only half of the data in $\T(1)\times \T(1)$ is needed to get the maximal surface geometric initial
 data. The other half of the phase space coordinates determines a complex structure on the maximal surface, which may or not coincide with the complex structure of the isothermal
 complex coordinate on this surface. This non-geometric half of the initial data can be interpreted as determining how the maximal surface is foliated by $|w|=const$ curves while
 the other, geometric half, determines the curve along which the maximal surface intersects the boundary at infinity, see Fig. 1. We have also seen that an equally good description
 of the same class of spacetimes is provided by $T^\ast \T(1)$, and that the generalized Mess map between the two descriptions is a symplectomorphism.

We have then studied the relation between the maximal surface description of AdS${}_3$ spacetimes given in this paper and the more standard holographic description by the
 Fefferman-Graham expansion of the spacetime metrics (\ref{metric-hol}). We have only been able to give an infinitesimal relation between two such metrics that are close to
 the standard metric on AdS${}_3$. However, the interpretation of such relation is a natural one. Namely, we interpret the phase space of AdS${}_3$ spacetimes as a deformation
 space of a given fixed reference spacetime. On one hand, in the usual holographic description we deform each asymptotic region of the reference spacetime with the group of
 asymptotic symmetries generated by nontrivial Brown-Henneaux vector fields. On the other, with the maximal surface parametrization, we consider quasiconformal deformations
 of the associate pair of hyperbolic surfaces obtained via the generalized Mess map. These are generated by harmonic Beltrami coefficients and the relation (\ref{rel}) is simply
 identifying these generators with the Brown-Henneaux vector fields. It is thus expected that such relation between generators can be extended to finite transformations thus
 identifying, to a certain extent, the asymptotic and quasiconformal deformation spaces. Note that this cannot be a one-to-one identification since, although enough to describe all
 possible asymptotically AdS${}_3$ metrics in a neighbourhood of conformal infinity, the Brown-Henneaux vector field does not contain the bulk moduli, that is, they do not fix the
 internal spacetime topology. This is the main advantage of the new parametrization proposed in this work as the maximal surface data also provides such moduli.

Even without a general finite relation between the descriptions, we were able to obtain an expression (\ref{charges}) for the charges (mass and angular momentum) of a spacetime
 in terms of data on the maximal surface. This expression admits an immediate generalization to the finite case, where the charges would simply be given by the real parts of the
 periods of the holomorphic quadratic differentials arising from the Bers embedding of $\T(1)\times \T(1)$. It would be very interesting to see that this is indeed the case for a
 general metric from our family. We leave this to future work. We have also shown that our description in terms of $\T(1)\times \T(1)$ is natural in terms of the Chern-Simons
 description of AdS${}_3$ gravity, in that the two Chern-Simons connections corresponding to our AdS${}_3$ metrics decouple with each being parametrized by a single copy of $\T(1)$.

The natural question that arises is what our constructions can add to the debate as to the microscopic origin of the entropy of 2+1 dimensional black holes. Here we can only give
 some speculations on this issue. As we have already mentioned in the Introduction, it seems sensible to approach the problem of quantum gravity in 2+1 dimensions as the problem of
 quantization of the moduli space of 2+1 dimensional constant curvature manifolds. In the context of negative cosmological constant all fixed spatial topology moduli spaces are
 realized as submanifolds of the universal moduli space described in the present work. The universal space therefore includes all possible multi-black-holes, together with the
 Brown-Henneaux excitations in each of their asymptotic regions. It also includes all compact spatial slice spacetimes (in this case one should simply take the initial data to be
 invariant under a Fuchsian group of a compact surface), but these spacetimes are unlikely to be relevant to the problem of BH entropy. One can then reformulate the question of
 computing the BH entropy as that of computing the partition function over all possible multi-black-hole spacetimes with fixed mass and angular momentum of one of the asymptotic
 regions. The entropy could then be extracted from this ``canonical'' partition function by the standard thermodynamic formulas. Our (infinitesimal case) expression (\ref{charges})
 for the charges is then the first step in this direction.

It would be very interesting if it were possible to reformulate the partition function computation as that in the context of some conformal field theory. In this respect we note that
 the Gauss-Codazzi equations that arise on the maximal surface in AdS${}_3$ are integrable, and are those of the so-called ${\rm sl}_2$ affine Toda system. It thus could be that the
 conformal field theory associated to the ${\rm sl}_2$ affine Toda is the CFT relevant for the quantum description of AdS${}_3$ gravity. We note that this CFT would naturally live on
 the maximal surface, not on the asymptotic boundary. But we have seen that the analytic continuation (to the imaginary time) of the functions on the AdS${}_3$ boundary cylinder has a
 natural interpretation in terms of data on the maximal surface. Thus, it appears that the Euclidean signature CFT on the spatial slice can, when analytically continued, be relevant
 for the AdS/CFT type description of 2+1 dimensional quantum gravity. Whether any of these speculations have a chance to come out true only future works on the subject can tell.

\section*{Acknowledgements}
KK was supported by a fellowship from the Alexander von Humboldt Foundation, Germany. The hospitality of the Albert Einstein Institute, Golm, is gratefully acknowledged. CS was
 supported by a University of Nottingham School of Mathematical Sciences Research Scholarship. The authors would like to thank Jorma Louko for comments on an earlier version of
 this work, and Catherine Meusburger for discussions on these topics. CS would also like to thank Luiz Fernando Carlvalho da Rocha for discussions.

\section{Appendix A: Universal Teichm\"uller theory}
We give a very basic introduction to the theory of universal Teichm\"uller space trying to keep the work as self-contained as possible. Our presentation follows closely the
 presentations of \cite{Ahlfors:2006,Gardiner:2000,Nag:1993,Takhtajan:2006}.

For a compact Riemann surface $\Sigma$ the Teichm\"uller space $\T(\Sigma)$ is defined as the space of conformal structures on $\Sigma$ modulo (small) diffeomorphisms in the
 connected component of the identity. This definition is usually described in an equivalent manner in terms of the possible hyperbolic structures on $\Sigma$. However, in
 generalizing the construction of Teichm\"uller space for noncompact Riemann surfaces with hyperbolic ends, it is important to keep track of the relation between the structures
 on the interior of the surface and its conformal boundary. The good definition of $\T(\Sigma)$ is then given in terms of quasiconformal deformations of the conformal structure
 of $\Sigma$. More concretely, Teichm\"uller space is then defined as the space of Beltrami coefficients on $\Sigma$ up to equivalence relation describing when two Beltrami
 coefficients define the same conformal structure.

To define the universal Teichm\"uller space, let $\Delta=\{w\in\hat\C;|z|<1\}$ and $\Delta^\ast=\{w\in\hat\C;| z|>1\}$ be the unit disc and its exterior in the Riemann sphere
 $\hat{\C}$ and let
\be\nonumber
L^\infty(\Delta)_1=\left\{\mu:\Delta\rightarrow\C;|\mu|_\infty=\sup_{\Delta}|\mu(w)|<1\right\},
\ee
be the unit ball in the space of bounded Beltrami differentials on $\Delta$. We define $\T(1)$ as the space of equivalence classes of such bounded Beltrami coefficients on $\Delta$,
\be\nonumber
\T(1)=L^\infty(\Delta)_1/\sim,
\ee
the equivalence relation being defined as follows.
{\flushleft\bf Model A.} Given two bounded Beltrami coefficients $\mu,\nu\in L^\infty(\Delta)_1$ one solves Beltrami equations in $\C$ with coefficients extended to $\Delta^\ast$ by
 reflection
\be\nonumber
\tilde\mu(w)=\begin{cases}\overline{\mu(1/\bar w)}w^2/\bar w^2,&w\in\Delta^\ast,\cr
\mu(w),& w\in\Delta,\end{cases}
\ee
similarly for $\nu$. Then $\mu,\nu$ are taken to be equivalent if the corresponding solutions, normalized to fix $-1$, $-\im$ and $1$, agree in $\bbS^1$
\be\nonumber
z_\mu\big|_{\bbS^1}=z_\nu\big|_{\bbS^1}.
\ee

{\flushleft\bf Model B.} Equivalently, one can define the equivalence relation by solving the Beltrami equations in $\C$ with Beltrami coefficients give by 
\be\nonumber
\tilde\mu(w)=\begin{cases}0,& w\in\Delta^\ast,\cr
\mu(w),&w\in\Delta,\end{cases}
\ee
similarly for $\nu$. Now, $\mu,\nu$ are considered equivalent if the corresponding solutions, normalized to have a simple pole of residue $1$ at $\infty$ and to satisfy
 $z(w)-w\rightarrow0$ for $w\rightarrow\infty$, agree on $\Delta^\ast$
\be\nonumber
z^\mu\big|_{\Delta^\ast}=z^\nu\big|_{\Delta^\ast}.
\ee

The equivalence relations above describe the conformal equivalence classes among the quasiconformal deformations of the conformal structure on $\Delta$. One can, therefore, describe
 universal Teichm\"uller space either as the space of (normalized) quasisymmetric homeomorphisms on $\bbS^1$ or the space of (normalized) univalent functions on $\Delta^\ast$.

Model B allows an important realization of the Teichm\"uller space as an embedded subspace of holomorphic quadratic differentials on $\Delta^\ast$
\be\nonumber
A_\infty(\Delta^\ast)=\{h:\Delta^\ast\rightarrow\C\textnormal{ holomorphic};|h(w)(1-|w|^2)^2|_\infty<\infty\}.
\ee
This is the so-called Bers embedding of $\T(1)$ and is obtained via Schwarzian derivative of $z^\mu|_{\Delta^\ast}$
\be\nonumber
\{z^\mu|_{\Delta^\ast},w\}=\frac{\partial_w\partial_w\partial_wz^\mu}{\partial_wz^\mu}-\frac{3}{2}\left(\frac{\partial_w\partial_wz^\mu}{\partial_wz^\mu}\right)^2.
\ee
This defines, in particular, a structure of complex Banach manifold on $\T(1)$, compatible with the one coming from $L^\infty(\Delta)_1$ via projection.

The space of bounded Beltrami coefficients $L^\infty(\Delta)_1$ also carries a group structure induced by the composition of quasiconformal maps. The group multiplication is
 defined as $\lambda=\nu\ast\mu$ iff the following relation is satisfied:
\be\nonumber
(\nu\circ z_\mu)=\frac{\lambda-\mu}{1-\lambda\bar\mu}\frac{\partial_wz_\mu}{\partial_{\bar w}\bar z_\mu}.
\ee
More explicitly, $\lambda$ is the Beltrami coefficient of $z_\lambda=z_\nu\circ z_\mu$ and is given by
\be\nonumber
\lambda=\frac{\mu+\nu\circ z_\mu\frac{\partial_{\bar w}\bar z_\mu}{\partial_wz_\mu}}{1+\bar\mu\nu\circ z_\mu\frac{\partial_{\bar w}\bar z_\mu}{\partial_wz_\mu}}.
\ee
Such group structure also descends to $\T(1)$.

\subsection*{Tangent Space}
Let's denote by $\Phi:L^\infty(\Delta)_1\to\T(1)$ the quotient map sending each bounded Beltrami coefficient $\mu$ on $\Delta$ into its equivalence class $[\mu]\in\T(1)$. Then,
 the derivative map $D_{\mu}\Phi:L^\infty(\Delta)\to T_{[\mu]}\T(1)$ identifies the tangent space to universal Teichm\"uller space $T_{[\mu]}\T(1)$ with the quotient space
 $L^\infty(z_\mu(\Delta))/N(z_\mu(\Delta))$ of the space of Beltrami coefficients on $z_\mu(\Delta)$ by its subspace $N(\Delta_\mu)$ of infinitesimally trivial coefficients, the
 kernel of $D_{\mu}\Phi$.

Let's set $\mu=0$ and consider the tangent space at base point of $\T(1)$. We denote by $\delta\mu$ an element of $L^\infty(\Delta)$ thought as a tangent vector at the origin. The
 infinitesimal version of Beltrami equation is then given by
$$\partial_{\bar w}f=t\delta\mu\partial_wf,$$
and its infinitesimal solutions can be written
$$f_{t\delta\mu}(w)=w+t\delta z+O(t^2),\qquad\partial_{\bar w}\delta z=\delta\mu.$$
Thus, a tangent vector $\delta\mu\in L^\infty(\Delta)$ defines a one parameter family of quasiconformal transformations from the A model procedure. We shall say that $\delta\mu$ is
 infinitesimally trivial if, to first order in $t$, the restriction to $\bbS^1$ of this family is given just by the identity transformation, that is, if the corresponding variation
 $\delta z|_{\bbS^1}$ vanishes identically. This, of course, means that $\delta\mu$ does not change the conformal structure on $\Delta$ and, therefore, that it is in the kernel
 $N(\Delta)$ of the derivative map $D_0\Phi:L^\infty(\Delta)\to T_{[0]}\T(1)$.

The infinitesimally trivial condition can be given different characterizations. It can be shown, see \cite{Ahlfors:2006}, that $\delta\mu\in N(\Delta)$ is equivalent to
$$\int_\Delta d^2w\,h\delta\mu=0,$$
for any holomorphic quadratic differential $h\in A_\infty(\Delta)$. Here, we define the space of holomorphic quadratic differentials as
\be\nonumber
A_\infty(\Delta)=\left\{h:\Delta\rightarrow\C\textnormal{  holomorphic};|h(w)(1-|w|^2)^2|_\infty<\infty\right\}
\ee
and we may write
$$N(\Delta)=\left\{\delta\mu\in L^\infty(\Delta);\int_\Delta d^2w\,h\delta\mu=0,\forall h\in A_{\infty}(\Delta)\right\}.$$

The map $D_0\Phi:L^\infty(\Delta)\to T_{[0]}\T(1)$ thus identifies the tangent space $T_{[0]}\T(1)$ with the space of harmonic Beltrami differentials on $\Delta$
\be\nonumber
\Omega^{-1,1}(\Delta)=\left\{\delta\mu=-\frac{(1-|w|^2)^2}{2}\overline{h(w)};h\in A_{\infty}(\Delta)\right\},
\ee
since the space $L^\infty(\Delta)$ can be decomposed as
\be\nonumber
L^\infty(\Delta)=N(\Delta)\oplus\Omega^{-1,1}(\Delta).
\ee

One may also think of the family $f_{t\delta\mu}$ as the one-parameter flow of the vector field $\delta z\partial_w$. Its restriction to $\bbS^1$ is
\be\nonumber
\delta z\partial_w\big|_{\bbS^1}=u(e^{\im\theta})\partial_\theta
\ee
with
\be\nonumber
u=\frac{\delta z(e^{\im\theta})}{\im e^{\im\theta}}=\sum_{k\neq-1,0,1}u_ke^{\im k\theta}.
\ee
This is an element of the so-called Zygmund class on $\bbS^1$, $\Lambda(\bbS^1)$, defined by
\be\nonumber
\Lambda(\bbS^1)=\left\{u:\bbS^1\rightarrow\R\textnormal{ continuous such that }A_u\in\Lambda(\R)\right\}
\ee
where
$A_u(x)=\frac{1}{2}(x^2+1)u\left(\frac{x-\im}{x+\im}\right)$ and
\be\nonumber
\Lambda(\R)=\{A:\R\rightarrow\R\textnormal{ continuous such that }\hspace{3cm}
\cr
| A(x+t)+A(x-t)-2A(x)|\leq \kappa| t|,\kappa>0\}.
\ee
Note that the coefficients  $u_{-1},u_0,u_1$ where dropped due to the normalization condition. Consequently, $u$ belongs to the quotient $\Lambda(\bbS^1)/{\rm M\ddot{o}b}(\bbS^1)$
 and the construction above provide an identification between $T_{[0]}\T(1)$ and the M\"obious normalized Zygmund class on $\bbS^1$.

The tangent space to the B-model universal Teichm\"uller space is obtained similarly by considering the infinitesimal solutions of Beltrami equation. Now, the one-parameter family
 $f^{t\mu}$ is of the form $$f^{t\mu}(w)=w+t\delta z(w)+O(t^2),$$ with the function $\delta z$ being holomorphic on $\Delta^\ast$. It thus admits an expansion in $\Delta^\ast$
\be\nonumber
f^{t\mu}(w)=w\Big(1+\frac{tc_2}{w^2}+\frac{tc_3}{w^3}+\ldots\Big).
\ee
The associated Bers embedding holomorphic quadratic differential can also be expanded in $\Delta^\ast$ as
\be\nonumber
h(w)=\frac{1}{w^4}\left(h_0+\frac{h_1}{w}+\frac{h_2}{w^2}+\ldots\right),
\ee
where the coefficients $h_k$ are related to those of $\delta z$ via
\be\nonumber
h_{k-2}=c_k(k-k^3),\qquad k\geq 2.
\ee
The relation between the coefficients in the A and B models now becomes quite simple
\be\nonumber
u_k=\im\bar c_k =\im\frac{\bar h_{k-2}}{(k-k^3)}, \quad k\geq 2,
\ee
see \cite{Nag:1993} for a proof. This could in fact be expected from the identification of $T_{[0]}\T(1)$ with the space of harmonic Beltrami coefficient $\Omega^{-1,1}(\Delta)$,
 in which the infinitesimal Beltrami coefficient is given in terms of a dual holomorphic quadratic differential $q\in A_{\infty}(\Delta)$. Writing 
\be\nonumber
q(w)=\sum_{k\geq0}q_{k}w^{k}
\ee
for the Laurent expansion of $q$, one can explicitly find the A model $\delta z$ by integration. For $w\in\Delta$ we get
\begin{align}\nonumber
\delta z(w)&=\frac{1}{2}\sum_{k\geq 2}\frac{\bar q_{k-2}\bar w^{k-1}}{(k-k^3)}\left[k(k+1)-2(k^2-1)|w|^2+k(k-1)|w|^4\right]+F(w),
\end{align}
where $F$ is some holomorphic function on $\Delta$, and, by reflection symmetry
$$\delta z(w)=-w^2\overline{\delta z(1/\bar w)},$$
for $w\in\Delta^\ast$
\begin{align}\nonumber
\delta z(w)=-\frac{1}{2}\sum_{k\geq 2}\frac{q_{k-2}\bar w^{-k-1}}{(k-k^3)}\left[k(k+1)|w|^4-2(k^2-1)|w|^2+k(k-1)\right]-w^2\overline{F(1/\bar w)}.
\end{align}
We may then write $F$ as
\be\nonumber
F(w)=\sum_{k\geq 0}v_kw^k
\ee
and restrict $\delta z$ to $\bbS^1$ to get
\begin{align}\nonumber
\delta z(e^{\im\theta})&=\sum_{k\geq 2}\frac{\bar q_{k-2}e^{-(k-1)\im\theta}}{(k-k^3)}+v_{0}+v_{1}e^{\im\theta}+v_{2}e^{2\im\theta}+\sum_{k\geq 2}v_{k+1}e^{(k+1)\im\theta}
\\  \nonumber
&=-\sum_{k\geq 2}\frac{q_{k-2}e^{(k+1)\im\theta}}{(k-k^3)}-\bar v_0e^{2\im\theta}-\bar v_1e^{\im\theta}-\bar v_2-\sum_{k\geq 2}\bar v_{k+1}e^{-(k-1)\im\theta}
\end{align}
so $v_0=-\bar v_2$, $v_1=-\bar v_1$, $v_{k+1}=\im u_k$ for $k\geq 2$. Thus
\begin{align}\nonumber
\delta z(w)&=\frac{1}{2}\sum_{k\geq 2}\frac{\bar q_{k-2}\bar w^{k-1}}{(k-k^3)}\left[k(k+1)-2(k^2-1)|w|^2+k(k-1)|w|^4\right]
\cr
&\hspace{1cm}+v_0+v_1w+v_2w^2-\sum_{k\geq 2}\frac{q_{k-2}w^{k+1}}{(k-k^3)},\quad w\in\Delta.
\end{align}
Note that, because of the normalization condition imposing $\delta z$ to vanish at $-1$, $-\im$ and $1$, the coefficients $v_0,v_1,v_2$ are completely determined by the
 $u_k$, $k\geq 2$. In fact, the M\"obious group is realized exactly as
\be\nonumber
{\rm M\ddot ob}(\bbS^1)\approx\left\{u(e^{\im\theta})=\bar u_{1}e^{-\im\theta}+u_0+u_1e^{\im\theta}\right\}
\ee
and, therefore, those coefficients are gauge. From now on we will drop the coefficients, understanding that they acquire the necessary values to make $\delta z$ vanish at
 $-1,-\im,1$.

We can now read off the Fourier coefficients of the Zygmund function $u$. With our choice of coefficients for $q$ as above, $u$ is simply given by
$$u(e^{\im\theta})=\sum_{k\neq-1,0,1}u_ke^{\im k\theta},$$
with $$u_k=\im\frac{q_{k-2}}{(k-k^3)},\qquad u_{-k}=\bar u_k$$In particular, the dual quadratic differential to $\delta\mu$ relates to the Bers embedding quadratic differential by
 the simple reflection rule
\be\nonumber
q(w)\in A_\infty(\Delta)\longmapsto h(w)=\overline{q(1/\bar w)}\frac{1}{w^4}\in A_\infty(\Delta^\ast).
\ee

\subsection*{Weil-Petersson hermitian metric}
The almost complex structure at the origin of $\T(1)$ is most clear from the model B point of view in which $J:T_{[0]}\T(1)\rightarrow T_{[0]}\T(1)$ is just
\be\nonumber
Jh=\im h.
\ee
By the isomorphism above described we have the almost complex structure
\be\nonumber
Ju=\im\sum_{k\neq -1,0,1}{\rm sgn}(k)u_ke^{\im n\theta}
\ee
on the space of normalized Zygmund class functions.

The Weil-Petersson hermitian metric on Teichm\"uller space of Riemann surfaces can be easily generalized to a (formal) hermitian metric on universal Teichm\"uller space. Explicitly, given $\delta\mu,\delta\nu\in T_{[0]}\T(1)$
we define
\be\nonumber
\langle\delta\mu,\delta\nu\rangle_{\rm WP}=\int_{\Delta}\frac{4d^2w}{(1-|w|^2)^2}\delta\mu(w)\delta\bar\nu(w).
\ee
Bers embedding then gives
\be\nonumber
\langle h,q\rangle_{\rm WP}=\int_{\Delta}d^2w(1-|w|^2)^2h(w)\bar q(w),
\ee
for $h,q\in A_\infty(\Delta^\ast)$, and the isomorphism above $\Lambda(\bbS^1)/{\rm M\ddot{o}b}(\bbS^1)\rightarrow A_{\infty}(\Delta^\ast)$
\be\nonumber
\langle u,v\rangle_{\rm WP}=\sum_{k,l\geq 2}k(k^2-1)l(l^2-1)u_k\bar v_l\int_{\Delta}d^2w(1-|w|^2)^2w^{k-2}\bar w^{l-2}.
\ee
Using
\be\nonumber
\int_{\Delta}d^2w(1-|w|^2)^2w^{k-2}\bar w^{l-2}=\int_\Delta drd\theta (1-r^2)^2r^{k+l-3}e^{\im(k-l)\theta}=\frac{2\pi}{k(k^2-k)}\delta_{kl}
\ee
we get the Weil-Petersson symplectic structure in terms of the coefficients of the Zygmund functions,
\be\nonumber
\langle u,v\rangle_{\rm WP}=2\pi\sum_{k\geq 2}k(k^2-1)u_k\bar v_l.
\ee

\end{document}